\documentclass[prd,showpacs,a4paper,draft,amsfonts,amssymb,amsmath,nofootinbib]{revtex4}
\newcommand{\dd}{\textrm{d\!I}}
\newcommand{\ww}{\wedge \kern-.75em \wedge}
\begin{document}

\title{Kinetic Terms for 2-Forms in Four Dimensions}

\author{J. Fernando \surname{Barbero G.}}
\email[]{jfbarbero@imaff.cfmac.csic.es} \affiliation{Instituto de
Matem\'aticas y F\'{\i}sica Fundamental, C.S.I.C., C/ Serrano
113bis, 28006 Madrid, Spain}
\author{Eduardo J. \surname{S. Villase\~nor}}
\email[]{eduardo@imaff.cfmac.csic.es} \affiliation{Instituto de
Matem\'aticas y F\'{\i}sica Fundamental, C.S.I.C., C/ Serrano
113bis, 28006 Madrid, Spain}

\date{May 8, 2002}

\begin{abstract}
We study the general form of the possible kinetic terms for 2-form
fields in four dimensions, under the restriction that they have a
semibounded energy density. This is done by using covariant
symplectic techniques and generalizes previous partial results in
this direction.
\end{abstract}

\pacs{11.15.-q, 03.65.Ca, 04.20.Fy}

\maketitle

\section{\label{Intro}Introduction}

A cursory look at the Quantum Field Theory literature may give the
impression that free (i.e. quadratic) Lagrangians are trivial
first steps in the road to the physically interesting interacting
theories. Once the important concepts of quantization in Fock
spaces, the definition of particle states, quantum numbers and so
on are understood, the emphasis shifts to the complicated problems
of dealing with interactions and computing cross sections. The
usual actions for the scalar, vector and spin $1/2$ fields seem to
cover all the possibilities as far as free theories are
concerned\footnote{Gravity can be derived from the quadratic
Fierz-Pauli Lagrangian.} so one might think that there is little
more to be learned about free actions. This, however, is not the
case. Although Poincar\'e invariance in Minkowski spacetime
certainly constraints the physical properties of the states
described by a given, \emph{arbitrarily general}, quadratic action
its precise particle content is not always obvious because it is
usually not evident how to ``diagonalize" and write it in terms of
the familiar free actions mentioned above. It is thus necessary to
find a way to classify the possible kinetic terms according to
their particle contents and their physical consistency (in
particular it is very important to know if their energy is
semibounded or not). In two previous papers \cite{eduferdif},
\cite{edufersform} we have obtained some partial results in this
direction. In \cite{eduferdif} we classified all the possible
diff-invariant terms in four dimensions. One of the interesting
results of that paper was finding out that quadratic
diff-invariant Lagrangians do not describe any local physical
degrees of freedom. In \cite{edufersform} we studied a family of
quadratic actions for $s$-form fields in Minkowski
spacetime\footnote{To make this paper as self-contained as
possible, we give a quick review of the results of that paper in
appendix \ref{R=0}.} given by
\begin{equation}
S_s[A]=\int_{\mathbb{R}^4}\left[dA^t \wedge *PdA+\delta A^t \wedge
*Q\delta A \right]\quad,\label{001}
\end{equation}
where $A$ is a set of $N$ $s$-form fields (with its transpose
denoted as $A^t$), $d$ is the exterior derivative, $*$ is the
Hodge dual used to define the dual derivative $\delta$  and
$\wedge$ the exterior product. $P$ and $Q$ are quadratic forms
represented by symmetric, real, $N\times N$ matrices. We found out
that all the Lagrangians of this type, with semibounded energy,
could be rewritten as the sum of free $s$-form Maxwell-like
actions and $(\partial_a A^a)^2$ terms for disjoint sets of
fields. Although (\ref{001}) describes a large family of quadratic
actions it is by no means the most general one. First of all, it
only involves bosonic $s$-form fields and not other types of
geometrical objects that may be important in some instances (for
example, twice covariant symmetric tensors); also, it does not
have cross terms involving forms of different type (no coupling
between, say, 1-forms and 2-forms). Finally, in the special case
of 2-forms, there is the possibility of adding an extra term,
involving a new ``coupling matrix" $R$, that does not exist in the
other cases. The purpose of this paper is to study the inclusion
of this additional term and develop the mathematical tools
necessary to complete the classification of all the possible
kinetic terms for 2-form fields in a four dimensional Minkowskian
background.

There are several reasons to consider this problem. First,
antisymmetric fields appear in physically relevant situations; it
is known, for example, that string theories contain states that
are described by this type of objects at low energies. They have
also been used in several proposed mass generation mechanisms for
gauge bosons without the introduction of Higgs fields (see
\cite{Harikumar:2001eb} and references therein). Second, we
believe that it is important to know all the free actions of a
certain type and consider the consistent introduction of
interactions along the lines of \cite{Henn} to explore the
possible existence of new models that can be studied by the usual
methods of perturbative Quantum Field Theory. Third, it is a first
step towards the classification of kinetic terms that couple
1-forms and 2-forms and their interactions (itself an open problem
\cite{Henneaux:1997ha}). Finally, we think that it is very
important to understand the physical content described by
\emph{general} free actions. In particular we want to know if,
after imposing the condition that the energy be semi-definite, it
is possible to describe particles with helicities different from
zero. If helicity 1 modes are present, then we could have
consistent actions for massless spin 1 fields maybe different from
the usual Yang-Mills ones. If we could find helicity 2 states, we
would have an alternative to the Fierz-Pauli Lagrangian that might
lead to a reasonable perturbative quantum field theory of gravity
after the introduction of self-interactions and matter couplings.
The results of \cite{edufersform} show that the action (\ref{001})
for 2-forms does not work in this sense; once we impose the
condition that the energy be semi-definite we find out that it
describes only massless scalars. However, as already quoted above,
for 2-forms there is the possibility of adding an extra term to
(\ref{001}). We investigate here if the previous conclusion
continues to apply to this case.

We study this problem within the covariant symplectic framework
introduced by Crnkovi\'c and Witten \cite{CR1, CR2, EW2} and used
by the authors in \cite{eduferdif,edufersform,Diffn}. The method
followed in those papers involved two main steps: the solution of
the field equations and the evaluation, on the solution space, of
the symplectic form derived from the action. The process of
solving the field equations depends on the geometrical setting of
the problem. In the case of the diff-invariant actions considered
in \cite{eduferdif,Diffn} they could be solved in a
straightforward way, however, in the presence of a Minkowskian
background \cite{edufersform} the process was more involved. It
consisted of three main steps: the identification of the relevant
algebraic sectors, the obtention of differential necessary
conditions and, finally, the resolution of the field equations.
The decomposition of the solution space that we found allowed us
to write down, in a very compact and convenient way the symplectic
structure on the solution space, the energy-momentum and the
helicity densities. We want to point out that it is possible to
follow more standard approaches, such as the familiar Dirac
analysis, to identify the physical degrees of freedom and gauge
symmetries of the problem that we consider. However, in order to
derive sufficient and necessary conditions to guarantee the
positiveness of energy in a generic case, we need to restrict the
energy to physical configurations (reduced phase space). This
requires, in practice, the resolution of the field equations and,
in our opinion, a covariant description provides the most elegant
and simple setting to study the problem.

The paper is structured as follows; after this introduction we
discuss the action, field equations and their solutions in section
\ref{actions...}. As we show it is possible to follow the same
steps and take advantage of the techniques developed in
\cite{edufersform} in the much more complicated case of the
general 2-form quadratic action that we study now. However, this
requires some additional steps, in particular we will have to
complexify the action, introduce suitable projection operators on
self-dual and anti-self dual 2-forms and recover the real theory
by implementing suitable reality conditions. In this respect the
program bears a striking resemblance with the original Ahstekar
variables formalism in which self-duality played and important
role \cite{Ashtekar:1987gu, Ashtekar:1986yd, Samuel:1987yy,
Jacobson:1988yy} and the use of reality conditions was a key
ingredient. In section \ref{symplectic} we obtain the symplectic
structure on the solution space that allows us to identify
canonical pairs of variables and the definition of Poisson
brackets. We use it also to derive the energy-momentum density and
the helicity of physical states in section \ref{Energy...}. We get
them as quadratic forms defined on the vector subspace defined by
the algebraic constraints obtained in the process of solving the
field equations. The main physical condition that we impose on the
models described by the family of actions considered in this paper
is the semi-boundedness of the energy, necessary at the classical
level to guarantee stability after the introduction of
interactions; this is discussed in detail in section
\ref{Semibound}. Once the physically consistent kinetic terms are
identified it is interesting to figure out if they can be written
in simple ``canonical" forms by linear field redefinitions. As
shown in \cite{edufersform} in the absence of the $R$-term it is
actually possible to write the kinetic term as a sum of ordinary
and ``dual" (of the $(\partial_a A^a)^2$ type) Maxwell
lagrangians; we show that this is also true here. Section
\ref{examples} is devoted to the study of the $R=0$ case as a
further check of the methods used in this paper. We show how to
recover the results obtained in previous work. Finally we end the
paper with our conclusions in section \ref{conclusions} and give
several appendices with detailed compilations of formulas and some
calculations needed in the main body of the paper.

\section{\label{actions...}Action, Field Equations, and Solutions}

We start by writing down the action (rather the family of actions)
that we consider in this paper
\begin{widetext}
\begin{eqnarray}
S_2[B]=\int_{\mathbb{R}^4}\left[\frac{1}{2}dB^t \wedge
*PdB+\frac{1}{2}\delta B^t \wedge *Q\delta B-dB^t\wedge R\delta B
\right]\quad,\label{002}
\end{eqnarray}
\end{widetext}
where $B$ denotes a set of $N$ real-valued 2-forms that we will
take as column vectors, $B^t$ is the transpose of $B$, $d$ is the
exterior differential and $\wedge$ is the exterior product defined
on $\mathbb{R}^4$. We have a Minkowski metric
$\eta_{ab}=\textrm{diag}(-+++)$ needed to define the Hodge dual
$*$ that we use to write down the adjoint exterior differential
$\delta$. We provide a dictionary with our conventions to
translate differential forms to the component framework in
appendix \ref{difforms}.  As in (\ref{001}), $P$ and $Q$ are real
$N\times N$ square symmetric matrices. The key new ingredient in
(\ref{002}) is the appearance of the last term (referred to in the
following as $R$-term) involving an arbitrary $N\times N$ square,
but not necesarily symmetric, real matrix. In four dimensions this
term can only be written for 2-forms. The Euler-Lagrange field
equations given by (\ref{002}) are
\begin{equation}
(P-*R^t)\delta d B+(Q-*R)d \delta B=0 \quad.\label{003}
\end{equation}
As in previous papers we start by solving this set of linear
partial differential equations. These equations involve a
combination of algebraic and differential operators. Whereas in
\cite{edufersform} the $P$ and $Q$ matrices commuted with the
differential operators $d$ and $\delta$, here the situation is
much more complicated because $P-*R^t$ and $Q-*R$ involve the
Hodge dual and, hence, one must be very careful with the order in
which these algebraic operators, $d$, and $\delta$ appear. In fact
this is the key difficulty that we have to overcome to treat this
problem along the lines followed for the action (\ref{001}).
Fortunately there is a clean way to solve this issue.

We will first consider Eq. (\ref{003}) for \emph{complex} $B$,
($B_{ab}(x)\in\mathbb{C}^N, \forall x\in\mathbb{R}^4$). The
linearity of the field equations and the fact that $P$, $Q$, and
$R$ are real objects tell us that both the real and imaginary part
of a complex solution are real solutions. Also, every complex
solution can be built from its real and imaginary parts.

We will introduce now the following set of projection operators
(acting on 2-forms):
\begin{eqnarray*}
 \pi_+&=&\frac{1}{2}(1+i*),\nonumber\\
 \pi_-&=&\frac{1}{2}(1-i*),\nonumber
 \end{eqnarray*}
that satisfy the following properties:
\begin{eqnarray*}
 & & \pi_+^2=\pi_+,\nonumber\\
 & & \pi_-^2=\pi_-,\nonumber\\
 & & \pi_-\pi_+=\pi_+\pi_-=0,\nonumber\\
 & & \pi_++\pi_-=1,\label{d}\\
 & & i(\pi_--\pi_+)=*\quad.\nonumber
\end{eqnarray*}
They also satisfy:
\begin{eqnarray*}
 \pi_+\delta d&=&\frac{1}{2}\square\pi_++\frac{1}{2}(\delta d-d\delta)\pi_-,\nonumber\\
 \pi_-\delta d&=&\frac{1}{2}\square\pi_-+\frac{1}{2}(\delta d-d\delta)\pi_+,\nonumber\\
 \pi_+d\delta&=&\frac{1}{2}\square\pi_+-\frac{1}{2}(\delta d-d\delta)\pi_-,\nonumber\\
 \pi_-d\delta&=&\frac{1}{2}\square\pi_--\frac{1}{2}(\delta d-d\delta)\pi_+.\nonumber
\end{eqnarray*}
We write now $B=(\pi_++\pi_-)B=\pi_+B+\pi_-B\equiv B_++B_-$ where
$\pi_+B_+=B_+$, $\pi_-B_-=B_-$, and $\pi_+B_-=\pi_-B_+=0$. It is
straightforward to see that $B$ is real if and only if
$B_-=\overline{B}_+$\footnote{In fact, if $B$ is real then
$\overline{B}_+=1/2(1-i*)\overline{B}=1/2(1-i*)B=B_-$. Conversely,
if $B_-=\overline{B}_+$ then $(1-i*)B=(1-i*)\overline{B}\,\,$. If
we write $B=\Re(B)+i \Im(B)$ the previous equality tells us that
$\Re(B)+*\Im(B)=\Re(B)-*\Im(B)$ and $\Im(B)-*
\Re(B)=-\Im(B)-*\Re(B)$ which implies $*\Im(B)=0 \Leftrightarrow
\Im(B)=0$.} (the overbar denotes complex conjugation). By writing
$P-*R^t=P(\pi_++\pi_-)-i(\pi_--\pi_+)R^t$ and
$Q-*R=Q(\pi_++\pi_-)-i(\pi_--\pi_+)R$  in the field equations and
projecting with $\pi_+$ and $\pi_-$ we see that they are
equivalent to:
\begin{eqnarray*}
 & &\pi_+\left[(P+iR^t)\delta dB+(Q+iR)d\delta B\right]=0,\nonumber\\
 & &\pi_-\left[(P-iR^t)\delta dB+(Q-iR)d\delta B\right]=0,\nonumber
\end{eqnarray*}
or
\begin{subequations}
 \label{008}
 \begin{eqnarray}
 & & (P+iR^t)\left[\square B_++(\delta
 d-d\delta)B_-\right]+(Q+iR)\left[\square B_+-(\delta
 d-d\delta)B_-\right]=0\qquad\label{008a}\\
 & & (P-iR^t)\left[\square B_-+(\delta
 d-d\delta)B_+\right]+(Q-iR)\left[\square B_--(\delta
 d-d\delta)B_+\right]=0\label{008b}.
 \end{eqnarray}
\end{subequations}
with $B_-=\overline{B}_+$. We see that by complexifying the action
and introducing the projection operators $\pi_+$ and $\pi_-$ we
have been able to write the field equations (\ref{003}) as
equations on the fields $B_+$ and $B_-$ \emph{that do not involve
the Hodge dual} $*$. This allows us to obtain the following
convenient necessary conditions by acting on (\ref{008}) with $d$
and $\delta$:
\begin{subequations}
 \label{009}
 \begin{eqnarray}
 & & \left[
   \begin{array}{rr}
    \,\, a & \quad b \\
    \,\, \overline{b} & \quad \overline{a}
   \end{array}\right]
 \left[
   \begin{array}{c}
    d\square B_+ \\
    d\square B_-
   \end{array}
 \right]=
 \left[
   \begin{array}{c}
    0 \\
    0
   \end{array}
 \right]\rightsquigarrow\mathcal{P}d\square\mathcal{B}=0\quad,\label{009a}\\
& &  \left[
   \begin{array}{rr}
    a & -b \\
    -\overline{b} & \overline{a}
   \end{array}\right]
 \left[
   \begin{array}{c}
    \delta \square B_+ \\
    \delta\square B_-
   \end{array}
 \right]=
 \left[
   \begin{array}{c}
    0 \\
    0
   \end{array}
 \right]\rightsquigarrow\mathcal{Q}\delta\square\mathcal{B}=0\quad,\label{009b}
 \end{eqnarray}
\end{subequations}
where $\mathcal{B}(x)\equiv\left[
   \begin{array}{c}
    B_+ \\
    B_-
   \end{array}
 \right]\in \mathbb{C}^{2N}$, $a\equiv P+Q+i(R+R^t)$,
and $b\equiv P-Q+i(R^t-R)$. The matrices $\mathcal{P}$ and
$\mathcal{Q}$ are \emph{symmetric} ($\mathcal{P}=\mathcal{P}^t$,
$\mathcal{Q}=\mathcal{Q}^t$) but, in general, \emph{not
hermitian}. The conditions (\ref{009}) have the same structure as
those found for the $R=0$ case in \cite{edufersform}.

We will discuss now several algebraic issues concerning the
matrices $\mathcal{P}$ and $\mathcal{Q}$. As already happened in
\cite{edufersform} it is not necessary to consider all the
possible quadratic forms $P$, $Q$ and $R$ because in some
situations it is trivially possible to eliminate some of the
fields appearing in the action by linear, non-singular field
redefinitions. The relevant condition to be imposed in this case
is
\begin{equation}
\ker(*P+R)\cap \ker(*Q+R^t)=\{0\} \label{010}
\end{equation}
with $*P+R$ and $*Q+R^t$ defined as linear operators in
$\Omega_2^N(\mathcal{M})$, the space of 2-form valued
$N$-dimensional vectors. The reason is that the field equations
(\ref{003}) can be rewritten in the form
$$d\delta(*P+R)B+\delta d(*Q+R^t)B=0;$$
suggesting that elements in $ \ker(*P+R)\cap \ker(*Q+R^t)$ should
play no role whatsoever. This turns out to be the case as can be
easily seen by rewriting the action (\ref{002}) as
\begin{eqnarray*}
S_2^{\prime}[B]=\frac{1}{2}\int_{\mathbb{R}^4}\left[\delta
B^t\wedge d (*Q+R^t)B-d B^t\wedge \delta(*P+R)B\right],
\end{eqnarray*} expanding $B=\rho^A
B_A+\sigma^{\overline{\alpha}} B_{\overline{\alpha}}$ (with
$(*P+R)\rho^A B_A=0$ and $(*Q+R^t) \rho^A B_A=0$ and
$\sigma^{\overline{\alpha}}$ in the complementary vector space)
and plugging this expression back in the previous form of the
action to check that all the $B^A$ fields drop out. The condition
(\ref{010}) can be reexpressed (see appendix \ref{alg}) as the
simple condition on $\mathcal{P}$ and $\mathcal{Q}$:
\begin{equation}
\ker\mathcal{P}\cap \ker\mathcal{Q}=\{0\}\quad.\label{011}
\end{equation}

We derive now some properties of $\ker\mathcal{P}$ and
$\ker\mathcal{Q}$. First it is straightforward to check that
$\left[\begin{array}{c} e_+ \\ e_- \end{array} \right]\in
\ker\mathcal{P}$ (with $e_+,\,
e_-\in \mathbb{C}^N$) if and only if $\left[\begin{array}{c} e_+ \\
-e_-
\end{array} \right]\in \ker\mathcal{Q}$. The main consequence of
this is the fact that
$$f:\ker\mathcal{P}\rightarrow \ker\mathcal{Q}:
\left[\begin{array}{c} e_+ \\ e_- \end{array} \right]\mapsto
\left[\begin{array}{c} e_+ \\ -e_-
\end{array} \right]$$ defines an isomorfism between $\ker\mathcal{P}$ and
$\ker\mathcal{Q}$ and hence
$\dim\ker\mathcal{P}=\dim\ker\mathcal{Q}\leq N$ as can be seen
from (\ref{011}). This isomorphism tells us how to get a basis of
$\ker\mathcal{Q}$ once we have a basis of $\ker\mathcal{P}$.
Second, every vector in a linear basis of $\ker\mathcal{P}$
can be written in the form  $\left[\begin{array}{c} v \\
\overline{v}
\end{array} \right]$ (for some $v\in\mathbb{C}^N$; see appendix \ref{alg}).
If $\{\left[\begin{array}{c} v_A \\
\overline{v}_A\end{array} \right]\}$ is a basis of
$\ker\mathcal{P}$ then $\{\left[\begin{array}{c} v_A \\
-\overline{v}_A\end{array} \right]\}$ is a basis of
$\ker\mathcal{Q}$. As $\ker\mathcal{P}\oplus
\ker\mathcal{Q}=Span\{
\left[\begin{array}{c} v_A \\ 0 \end{array}\right],\left[\begin{array}{c} 0 \\
\overline{v}_A\end{array} \right]\}$ we can complete this linearly
independent set to a basis of $\mathbb{C}^{2N}$ with $\{
\left[\begin{array}{c} e_{\alpha} \\
0 \end{array}\right],\left[\begin{array}{c} 0 \\
\overline{e}_{\alpha}\end{array} \right]\}$ such that $\{v_A,
e_{\alpha}\}$ are linearly independent as $\mathbb{C}^N$ vectors
and build the following convenient basis for $\mathbb{C}^{2N}$
$$\left\{\left[\begin{array}{c} v_A \\
\overline{v}_A\end{array} \right],\left[\begin{array}{c} v_A \\
-\overline{v}_A \end{array}\right],\left[\begin{array}{c} e_{\alpha} \\
\overline{e}_{\alpha}\end{array}\right],\left[\begin{array}{c} e_{\alpha} \\
-\overline{e}_{\alpha} \end{array}\right]\right\}.$$ We expand an
arbitrary vector $\mathcal{B}\in\mathbb{C}^{2N}$ in this basis as
$$
\mathcal{B}=B_p^A\left[\begin{array}{c} v_A \\
\overline{v}_A\end{array} \right]+B_q^A\left[\begin{array}{c} v_A \\
-\overline{v}_A\end{array} \right]+B_{\alpha}^+\left[\begin{array}{c} e_{\alpha} \\
\overline{e}_{\alpha} \end{array}\right]+
B_{\alpha}^-\left[\begin{array}{c} e_{\alpha} \\
-\overline{e}_{\alpha} \end{array}\right].$$ We use this expansion
to solve the necessary conditions (\ref{009}); in order to do this
we must enforce that $\pi_+B_-=\pi_-B_+=0$ in $\mathcal{B}$, which
are equivalent to the conditions $B^A_p=i*B^A_q$ and
$B_+^{\alpha}=i*B_-^{\alpha}$ so that, in terms of new fields
$\tilde{B}$ and $\hat{B}$ we have
\begin{eqnarray*}
& & B_+=(1+i*)[\tilde{B}^Av_A+\hat{B}^{\alpha}e_{\alpha}]\\
& &
B_-=(i*-1)[\tilde{B}^A\overline{v}_A+\hat{B}^{\alpha}\overline{e}_{\alpha}].
\end{eqnarray*}
We also have to impose the reality conditions; they simply tell us
that both $\tilde{B}^A$ and $\hat{B}_{\alpha}$ must be purely
imaginary. We conclude that $\mathcal{B}$ must have the form
\begin{equation}
\mathcal{B}=*B^A\left[\begin{array}{c} v_A \\
\overline{v}_A\end{array} \right]-iB^A\left[\begin{array}{c} v_A \\
-\overline{v}_A\end{array} \right]+b^{\alpha}\left[\begin{array}{c} e_{\alpha} \\
\overline{e}_{\alpha} \end{array}\right]+i*b^{\alpha}
\left[\begin{array}{c} e_{\alpha} \\
-\overline{e}_{\alpha} \end{array}\right] \label{012}
\end{equation} with
$B^A,\,b_{\alpha}$ real functions. Taking into account that the
vectors in both the following sets are linearly independent
$$\left\{\mathcal{P}\left[\begin{array}{c} v_A \\
-\overline{v}_A\end{array} \right],\,\mathcal{P}\left[\begin{array}{c} e_{\alpha} \\
\overline{e}_{\alpha}\end{array}\right],
\mathcal{P}\left[\begin{array}{c} e_{\alpha} \\
-\overline{e}_{\alpha} \end{array}\right]\right\},\quad
\left\{\mathcal{Q}\left[\begin{array}{c} v_A \\
\overline{v}_A\end{array} \right],\,
\mathcal{Q}\left[\begin{array}{c} e_{\alpha} \\
\overline{e}_{\alpha}\end{array}\right],
\mathcal{Q}\left[\begin{array}{c} e_{\alpha} \\
-\overline{e}_{\alpha} \end{array}\right]\right\}.$$
 The necessary conditions are simply
\begin{subequations}
\label{013}
\begin{eqnarray}
\square dB^A=0\label{013a}\\
\square db^{\alpha}=0\label{013b}\\
\square \delta b^{\alpha}=0\label{013c}
\end{eqnarray}
\end{subequations}
They have the same form as the ones found in \cite{edufersform} in
the absence of the $R$-term and can be easily solved by using the
procedure described in that paper. It should be pointed out that
one of the advantages of the method that we are using here is that
the structure of the necessary conditions is the same as in
previous, more simple, models \cite{edufersform}. The solutions to
(\ref{013}) are
\begin{subequations}
 \label{014}
 \begin{eqnarray}
 B^A=\gamma^A+d\Lambda^A,\label{014a}\\
 b^{\alpha}=\gamma^{\alpha}+d\delta\gamma^{\alpha}_{\scriptscriptstyle H},
 \label{014b}
 \end{eqnarray}
\end{subequations}
where $\Lambda^A$ is an arbitrary real function and $\square
\gamma^{\alpha}=0$, $\square \gamma^A=0$,
$\square^2\gamma^{\alpha}_{\scriptscriptstyle H}=0$.

To complete the solution to the field equations we plug
(\ref{014}) into the field equations (\ref{008}) to get
\begin{widetext}
\begin{equation}
(1+i*)\left[(ae_{\alpha}-b\overline{e}_{\alpha})d\delta d\delta
\gamma^{\alpha}_{\scriptscriptstyle H}+2i\delta d
\gamma^A(b\overline{v}_A)+2\delta
d\gamma^{\alpha}(b\overline{e}_{\alpha})\right]=0 \label{015}
\end{equation}
\end{widetext}
and its complex conjugate. In order to continue we proceed as in
\cite{edufersform} by introducing Fourier tranforms. To this end
we choose an inertial coordinate system $(\vec{x},t)$ in
$\mathbb{R}^4$ and define
\begin{eqnarray*}
f(\vec{x},t)=\frac{1}{(2\pi)^{3/2}}\int_{\mathbb{R}^3}\frac{d^3\vec{k}}{w}
f(\vec{k},t)e^{i\vec{k}\cdot \vec{x}} \quad;
\end{eqnarray*}
and its inverse
\begin{eqnarray*}
\frac{1}{w}f(\vec{k},t)=\frac{1}{(2\pi)^{3/2}}\int_{\mathbb{R}^3}d^3\vec{x}
f(\vec{x},t)e^{-i\vec{k}\cdot \vec{x}}
\end{eqnarray*}
(where we use 3-dimensional vector notation in $\mathbb{R}^3$ and
denote the usual Euclidean scalar product with a dot). We have
introduced $w=+\sqrt{\vec{k}\cdot \vec{k}}$ in the previous
definition in order to explicitly have a Lorentz invariant
measure. We use the same letter to represent a field and its
Fourier transform and for real fields we must have
$f(\vec{k},t)=\bar{f}(-\vec{k},t)$.  Using the formulas introduced
in appendix \ref{difforms}, in particular the definitions of + and
$-$ fields and $e_{ij}(\vec{k})$, (\ref{015}) may be written in
the two following alternative forms
\begin{widetext}
\begin{subequations}
 \label{016}
 \begin{eqnarray}
& & \left(\delta_{ij}+\frac{1}{3}we_{ij}(\vec{k})\right)
\left[\sigma^{\alpha}_j(\vec{k})(ae_{\alpha}-b\overline{e}_{\alpha})+2ib^A_j(\vec{k})
(a v_A)-2b^{\alpha}_j(\vec{k})(b\overline{e}_{\alpha})\right]=0 \label{016a}\\
& & \left(\delta_{ij}-\frac{1}{3}we_{ij}(\vec{k})\right)
\left[\sigma^{\alpha}_j(\vec{k})(\overline{a}\overline{e}_{\alpha}-
\overline{b}e_{\alpha})-2ib^A_j(\vec{k})(\overline{a}\overline{v}_A)-2b^{\alpha}_j(\vec{k})
(\overline{b}e_{\alpha})\right]=0
 \label{016b}
 \end{eqnarray}
\end{subequations}
\begin{subequations}
 \label{017}
 \begin{eqnarray}
\sigma^{\alpha}_-(\vec{k})(ae_{\alpha}-b\overline{e}_{\alpha})+2ib^A_-(\vec{k})
(a v_A)-2b^{\alpha}_-(\vec{k})(b\overline{e}_{\alpha})=0 \label{017a}\\
\sigma^{\alpha}_+(\vec{k})(\overline{a}\overline{e}_{\alpha}-
\overline{b}e_{\alpha})-2ib^A_+(\vec{k})
(\overline{a}\overline{v}_A)-2b^{\alpha}_+(\vec{k})
(\overline{b}e_{\alpha})=0
 \label{017b}
 \end{eqnarray}
\end{subequations}
\end{widetext}
where we have used $av_A+b\overline{v}_A=0$. The introduction of
the $+$ and $-$ fields is suggested, in part, by the simple form
that these algebraic conditions take and the fact that they
decouple in terms of them. Notice that, in each pair, the
conditions imposed on the field components are
\textit{independent} and also that the equations appearing in each
pair are \emph{not} complex conjugates of each other. In practice,
it is convenient to project the last two equations onto the bases
of $\mathbb{C}^N$ given, respectively, by $\{v_A, e_{\alpha}\}$
and $\{\bar{v}_A, \bar{e}_{\alpha}\}$ to get
\begin{widetext}
\begin{subequations}
 \label{0017}
 \begin{eqnarray}
(e^t_{\alpha}ae_{\beta}-e^t_{\alpha}b\bar{e}_{\beta})
\sigma^{\beta}_-(\vec{k})+2i(e^t_{\alpha}av_A)b^A_-(\vec{k})
-2(e^t_{\alpha}b\bar{e}_{\beta})b^{\beta}_-(\vec{k})=0 \label{0017a}\\
2\Re(v_A^tae_{\beta})\sigma^{\beta}_-(\vec{k})+
2i\Re(v_A^tav_B)b^B_-(\vec{k})+2(
\bar{v}_A^t\bar{a}\bar{e}_{\beta})b^{\beta}_-(\vec{k})=0\label{0017b}\\
(\bar{e}^t_{\alpha}\bar{a}\bar{e}_{\beta}-\bar{e}^t_{\alpha}\bar{b}e_{\beta})
\sigma^{\beta}_+(\vec{k})-2i(\bar{e}^t_{\alpha}\bar{a}\bar{v}_A)b^A_+(\vec{k})
-2(\bar{e}^t_{\alpha}\bar{b}e_{\beta})b^{\beta}_+(\vec{k})=0\label{0017c}\\
2\Re(v_A^tae_{\beta})\sigma^{\beta}_+(\vec{k})-
2i\Re(v_A^tav_B)b^B_+(\vec{k})+2(v_A^tae_{\beta})b^{\beta}_+(\vec{k})=0.\label{0017d}
 \end{eqnarray}
\end{subequations}
\end{widetext}
We will use them in this form to discuss the conditions leading to
a semibounded energy.

 This completes the resolution of the field
equations that we need in the next step to find the physical
degrees of freedom and gauge symmetries described by the action
(\ref{002}). A final comment related to the general structure of
the action and the resolution of the field equations is the
following. The action (\ref{002}) can be rewritten as
\begin{equation}
S_2=\frac{1}{4}\int_{\mathbb{R}^4}\left[d\mathcal{B}^t
\wedge*\mathcal{P}d\mathcal{B}+\delta\mathcal{B}^t
\wedge*\mathcal{Q}\delta\mathcal{B}\right] \label{018}
\end{equation}
with the additional condition $\pi_-B_+=\pi_+B_-=0$ implemented by
using suitable Lagrange multipliers $\lambda_+$,  $\lambda_-$ and
adding to (\ref{018}) the extra term
$\lambda_-\pi_-B_++\lambda_+\pi_+B_-$. In this form it is easy to
justify our choice of $\mathcal{P}$ and $\mathcal{Q}$ such that
$\ker\mathcal{P}\cap \ker\mathcal{Q}=\{0\}$ and take advantage of
the results already obtained in \cite{edufersform} in the absence
of the $R$-term. Also, we see that as $\pi_-B_+=\pi_+B_-=0$ and
the reality conditions do not involve time derivatives the
symplectic structure remains unchanged so, as we show in the
following, several interesting objects can be obtained from known
results for the $R=0$ case by simply restricting to the linear
subspaces defined by these conditions.

\section{\label{symplectic}Symplectic Structure, Gauge transformations,
and Physical Degrees of freedom}

Once we know the solutions to the field equations we can study the
physical content of our action (\ref{002}) by computing the
symplectic structure in the space of solutions to the field
equations. For a general Lagrangian $L(\xi, d\xi, \delta\xi)$ for
a s-form field $\xi$ we have that the symplectic form is
$$
\Omega(\xi)=\mathbb{\dd}\xi\ww \dd \frac{\partial L}{\partial
d\xi}+(-1)^{s+1}*\dd\xi\ww*\dd \frac{\partial L}{\partial
\delta\xi}
$$
where the derivatives of the Lagrangian are defined according to
\begin{eqnarray*}
L(\xi+\epsilon, d\xi+d\epsilon, \delta\xi+\delta\epsilon)&=&
L(\xi, d\xi, \delta\xi)+\epsilon\wedge\frac{\partial
L}{\partial\xi}(\xi, d\xi,\delta\xi)\\
& &+d\epsilon\wedge\frac{\partial L}{\partial d\xi}(\xi, d\xi,
\delta\xi) +\delta\epsilon\wedge\frac{\partial
L}{\partial\delta\xi}(\xi, d\xi, \delta\xi)+O(\epsilon^2)
\end{eqnarray*}
and $\dd$, $\wedge\!\!\!\wedge$ refer to the space of fields. From
the action (\ref{002}) we get
\begin{eqnarray*}
\Omega(B)=\dd B^t\ww*Pd\dd B+[Q\delta \dd B]^t\ww*d\dd B-\dd
B^t\!\ww R\delta\dd B-*\dd B^t\ww R^td \dd B.& &
\end{eqnarray*}
It is straightforward to check that $d\Omega(B)=0$ on solutions to
the field equations (notice that we are using here the spacetime
exterior derivative). This allows us to finally write the
symplectic structure in phase space as
\begin{eqnarray}
& & \Omega=\int_{\mathbb{R}^3}\left\{[\delta(*P+R)\dd B]^t\ww\dd B
+[*d(*Q+R^t)\dd B]^t\ww
*\dd B\right\}\label{020}\\
& & \hspace{.4cm}=\frac{1}{2}\int_{\mathbb{R}^3}\left[ \dd
\mathcal{B}^t\ww*\mathcal{P}d\dd\mathcal{B}-*\dd
\mathcal{B}^t\ww\mathcal{Q}\delta\dd \mathcal{B}\right].\nonumber
\end{eqnarray}
Plugging (\ref{012}) in (\ref{020}) we find
\begin{eqnarray}
\Omega=\!\int_{\mathbb{R}^3}\big\{4\dd B^A\!\ww*\Im_{A\alpha}d \dd
b^{\alpha}\!-4\dd B^A\!\ww*\Re_{AB}d\dd B^B +4\dd
B^A\ww\Re_{A\alpha}\delta\dd b^{\alpha}
& & \nonumber\\
+4\dd b^{\alpha}\ww*\Im_{\alpha A}d\dd B^A+2\dd
b^{\alpha}\ww*\Re^+_{\alpha\beta}d\dd b^{\beta}-2\dd
b^{\alpha}\ww\Im^-_{\alpha\beta}\delta\dd b^{\beta} & & \nonumber\\
 +4*\dd b^{\alpha}\ww*\Re_{\alpha A}d\dd B^A -2*\dd
b^{\alpha} \ww *\Im^+_{\alpha\beta}d \dd b^{\beta}-2*\dd
b^{\alpha} \ww\Re^-_{\alpha\beta}\delta \dd b^{\beta}\big\}&
&\nonumber
\end{eqnarray}
where  we have introduced the following notation:
\begin{eqnarray*}
\Re_{AB}\equiv \Re[v_A^tav_B]\quad (\Im[v_A^tav_B]=0)\nonumber\\
\Im_{A\alpha}\equiv\Im[v_A^tae_{\alpha}]\quad \Im_{\alpha
A }\equiv\Im[e_{\alpha}^tav_A]\nonumber\\
\Re_{A\alpha}\equiv\Re[v_A^tae_{\alpha}]\quad\Re_{\alpha
A } \equiv\Re[e_{\alpha}^tav_A]\nonumber\\
\Re^+_{\alpha\beta}\equiv\Re[e_{\alpha}^tae_{\beta}+e_{\alpha}^t
b\overline{e}_{\beta}]\nonumber\\
\Re^-_{\alpha\beta}\equiv\Re[e_{\alpha}^tae_{\beta}-e_{\alpha}^t
b\overline{e}_{\beta}]\nonumber\\
\Im^+_{\alpha\beta}\equiv\Im[e_{\alpha}^tae_{\beta}+e_{\alpha}^t
b\overline{e}_{\beta}]\nonumber\\
\Im^-_{\alpha\beta}\equiv\Im[e_{\alpha}^tae_{\beta}-e_{\alpha}^t
b\overline{e}_{\beta}]\nonumber\\
\Im^b_{\alpha\beta}\equiv\Im[e_{\alpha}^tb\overline{e}_{\beta}]\quad
\Im^a_{\alpha\beta}\equiv\Im[e_{\alpha}^ta\overline{e}_{\beta}]\nonumber\\
\Re^b_{\alpha\beta}\equiv\Re[e_{\alpha}^tb\overline{e}_{\beta}]\quad
\Re^a_{\alpha\beta}\equiv\Re[e_{\alpha}^ta\overline{e}_{\beta}].\nonumber
\end{eqnarray*}
It is important to remember that the derivative operators $d$ and
$\delta$ that appear in this expression are four dimensional and
the integral is over $\mathbb{R}^3$. In terms of Fourier modes the
symplectic structure is given by
\begin{eqnarray}
\Omega=\int_{\mathbb{R}^3}\frac{d^3k}{w}\bigg\{
16i\dd\overline{\beta}^A_{ij}(\vec{k})
\Re_{AB}\ww\dd\beta^B_{ij}(\vec{k})
+8i\dd\beta^{\alpha}_{ij}(\vec{k})\Re^+_{\alpha\beta}\ww\dd
\overline{\beta}^{\beta}_{ij}(\vec{k})& & \nonumber\\
+16i\left[\dd\beta^A_{ij}(\vec{k})\Im_{A\alpha}\ww\dd
\overline{\beta}^{\alpha}_{ij}(\vec{k})-\dd\overline{\beta}^A_{ij}(\vec{k})
\Im_{A\alpha}\ww\dd\beta^{\alpha}_{ij}(\vec{k})\right]& &
\nonumber\\
+\frac{16}{3}iwe_{ij}(\vec{k})\left[\dd\beta^A_{ij}(\vec{k})
\Re_{A\alpha}\ww \dd \overline{a}^{\alpha}(\vec{k})
+\dd\overline{\beta}^A_{ij}(\vec{k}) \Re_{A\alpha}\ww \dd
a^{\alpha}(\vec{k})\right]& &\nonumber\\
-\frac{8}{3}iwe_{ij}(\vec{k})\left[\dd\beta^{\alpha}_{ij}(\vec{k})
\Im^-_{\alpha\beta}\ww\dd\overline{a}^{\beta}(\vec{k})+
\dd\overline{\beta}^{\alpha}_{ij}(\vec{k})
\Im^-_{\alpha\beta}\ww\dd a^{\beta}(\vec{k})\right]& &\nonumber\\
-16i\dd a^{\alpha}(\vec{k})\Re^-_{\alpha\beta}
\ww\dd\overline{a}^{\beta}(\vec{k})+\frac{8}{3}we_{ij}(\vec{k})\dd\sigma^{\alpha}_i(\vec{k})
\Im^b_{\alpha\beta}\ww
\dd\overline{\sigma}^{\beta}_j(\vec{k})& &\nonumber\\
+\frac{4}{3}we_{ij}(\vec{k})\left[\dd\sigma^{\alpha}_{i}(\vec{k})
(\Im^+_{\alpha\beta}+2\Im^b_{\alpha\beta})
\ww\dd\overline{b}^{\beta}_j(\vec{k})-\dd\overline{\sigma}^{\alpha}_{i}(\vec{k})
(\Im^+_{\alpha\beta}+2\Im^b_{\alpha\beta})\ww\dd
b^{\beta}_j(\vec{k})\right]& &\nonumber\\
-4i\left[\dd\sigma^{\alpha}_{i}(\vec{k}) \Re^+_{\alpha\beta}
\ww\dd\overline{b}^{\beta}_i(\vec{k})-\dd\overline{\sigma}^{\alpha}_{i}(\vec{k})
\Re^+_{\alpha\beta}\ww\dd
b^{\beta}_i(\vec{k})\right]& &\nonumber\\
+\frac{16}{3}we_{ij}(\vec{k})\dd
b^{\alpha}_i(\vec{k})\Im^b_{\alpha\beta}\ww
\dd\overline{b}^{\beta}_{j}(\vec{k})-16i\dd
b^{\alpha}_i\Re^+_{\alpha\beta}\ww\dd\overline{b}^{\beta}_i(\vec{k})& &\nonumber\\
-8\left[\dd\alpha^{\alpha}_i(\vec{k})\Re^+_{\alpha\beta}
\ww\dd\overline{b}^{\beta}_i(\vec{k})+\dd\overline{\alpha}^{\alpha}_i(\vec{k})
\Re^+_{\alpha\beta}\ww\dd
b^{\beta}_i(\vec{k})\right]& &\nonumber\\
+\frac{8}{3}iwe_{ij}(\vec{k})\left[\dd\alpha^{\alpha}_i(\vec{k})
\Im^-_{\alpha\beta}
\ww\dd[\overline{b}^{\beta}_j(\vec{k})+\overline{\sigma}^{\beta}_j(\vec{k})]
+\dd\overline{\alpha}^{\alpha}_i(\vec{k}) \Im^-_{\alpha\beta}
\ww\dd [b^{\beta}_j(\vec{k})+\sigma^{\beta}_j(\vec{k})]\right]& &\nonumber\\
-8i [2\dd b^{\alpha}_i(\vec{k})+\dd
\sigma^{\alpha}_i(\vec{k})-2i\dd \alpha^{\alpha}_i(\vec{k})]
\Im_{\alpha A}\ww\dd\overline{b}^A_i(\vec{k}) &
&\nonumber\\+8i[2\dd \overline{b}^{\alpha}_i(\vec{k})+\dd
\overline{\sigma}^{\alpha}_i(\vec{k})+2i\dd
\overline{\alpha}^{\alpha}_i(\vec{k})] \Im_{\alpha A}\ww\dd
b^A_i(\vec{k})\bigg\}& &\nonumber
\end{eqnarray}
where we have made use of the algebraic conditions (\ref{016}) and
it must be remembered that this symplectic form is defined on the
solution space, i.e. when these conditions hold. Although its
structure is rather complicated some features are already evident.
For example, we can see that the arbitrary functions $\Lambda^A$
do not appear, this means that they do not represent any physical
degrees of freedom and should be taken as gauge parameters. In
order to see this we collect the terms involving $\Lambda^A$
\begin{eqnarray}
& &
4\int_{\mathbb{R}^3}d\dd\Lambda^A\ww*\biglb\{-\Re_{AB}d\dd\gamma^B+
\Re_{A\alpha}*\delta\dd\gamma^{\alpha}+\Re_{A\alpha} *\delta
d\delta \dd\gamma^{\alpha}_{\scriptscriptstyle
{H}}+\Im_{A\alpha}d\dd\gamma^{\alpha}\bigrb\}.\nonumber
\end{eqnarray}
The integrand in the previous expression can be written as
\begin{eqnarray}
& & d\biglb\{\dd\Lambda^A\ww*\biglb[-\Re_{AB}d\dd\gamma^B+
\Re_{A\alpha}*\delta\dd\gamma^{\alpha}+\Re_{A\alpha} *\delta
d\delta \dd\gamma^{\alpha}_{\scriptscriptstyle
{H}}+\Im_{A\alpha}d\dd\gamma^{\alpha}\bigrb]\bigrb\}\nonumber\\
& &-\dd\Lambda^A\ww*\biglb[-\Re_{AB}\delta d\dd\gamma^B+
\Re_{A\alpha}*d\delta\dd\gamma^{\alpha}+\Re_{A\alpha} *d\delta
d\delta \dd\gamma^{\alpha}_{\scriptscriptstyle
{H}}+\Im_{A\alpha}\delta d\dd\gamma^{\alpha}\biglb]\nonumber,
\end{eqnarray}
the first piece is an exact four dimensional 3-form and hence its
pull-back onto $\mathbb{R}^3$ is exact too so the integral over
$\mathbb{R}^3$ is zero. The second part is zero as a consequence
of the condition (\ref{015}); in fact, if one projects (\ref{015})
along the direction of $v_A$ and takes into account that
$\gamma^{\alpha}$ satisfies the wave equation one easily gets
\begin{eqnarray}
& &-\Re_{AB}\delta d\dd\gamma^B+
\Re_{A\alpha}*d\delta\dd\gamma^{\alpha}+\Re_{A\alpha} *d\delta
d\delta \dd\gamma^{\alpha}_{\scriptscriptstyle
{H}}+\Im_{A\alpha}\delta d\dd\gamma^{\alpha}=0\quad. \nonumber
\end{eqnarray}
The degrees of freedom for a specific choice of $P$, $Q$, and $R$
can be identified by solving the algebraic conditions in terms of
a set of independent fields and finding out which among them (and
the remaining fields that do not appear in these conditions)
appear in the symplectic structure.

\section{\label{Energy...}Energy and angular momentum}

The calculation of the energy can be performed \cite{edufersform}
by obtaining first the transformation of the Fourier components of
the fields under space-time translations, and using the fact that
if the symplectic form $\Omega$ is invariant under a group of
transformations locally we have $i_v\Omega=\dd H$, where $H$ is
the generator of the symmetry and $v$ is a tangent vector to an
orbit of this group. For a Poincar\'e invariant action this
procedure gives the energy-momentum density in a very convenient
way. The only difficulty, in practice, comes from the fact that
the algebraic conditions (\ref{016},\ref{017}) must be used in a
somewhat non obvious way to write $i_v\Omega$ as a exterior
derivative of something. In the present case there is an
alternative method that quickly leads to the correct answer. The
idea is to write $\mathcal{B}$ in terms of $\gamma^A$,
$\gamma^{\alpha}$, $\gamma^{\alpha}_{\scriptscriptstyle H}$ and
$\Lambda^A$ by plugging (\ref{014}) in (\ref{012}) and notice that
its structure is the same as the one obtained in
\cite{edufersform} for the solutions to the field equations if one
makes the identifications (splitting the $e_r$ sector in $+$ and
$-$ subsectors)
$$e^p\!\leftrightsquigarrow\!\!\left[\begin{array}{c} v_A \\
\bar{v}_A \end{array}\right]\!,\, e^q\!\leftrightsquigarrow\!\!
\left[\begin{array}{c} v_A \\-\bar{v}_A \end{array}\right]\!,\,
e^r_+\!\leftrightsquigarrow\!\! \left[\begin{array}{c} e_{\alpha} \\
\bar{e}_{\alpha} \end{array}\right]\!,\,
e^r_-\!\leftrightsquigarrow \!\!\left[\begin{array}{c} e_{\alpha} \\
-\bar{e}_{\alpha} \end{array}\right]\!.$$ The Fourier components
of the $A_p$, $A_q$, and $A_r$ fields (see appendix \ref{R=0}) can
be put in one to one correspondence with those of $\gamma^A$,
$\gamma^{\alpha}$, and $\gamma^{\alpha}_{\scriptscriptstyle H}$ as
follows
\begin{eqnarray*}
\begin{array}{ll}
\beta^q_{ij}(\vec{k})\rightsquigarrow -i
\beta^A_{ij}(\vec{k})\quad &
\bar{\beta}^q_{ij}(\vec{k})\rightsquigarrow -i
\bar{\beta}^A_{ij}(\vec{k})\\
b_i^q(\vec{k})\rightsquigarrow -i b^A_i(\vec{k})\quad &
\bar{b}_i^q(\vec{k})\rightsquigarrow -i \bar{b}^A_i(\vec{k})
\end{array}
\end{eqnarray*}
\begin{eqnarray*}
\begin{array}{ll}
\alpha^p_i(\vec{k})\rightsquigarrow
\frac{2}{3}we_{ij}(\vec{k})b^A_j\quad &
\bar{\alpha}^p_i(\vec{k})\rightsquigarrow
-\frac{2}{3}we_{ij}(\vec{k})\bar{b}^A_j\\
a^p(\vec{k})\rightsquigarrow-\frac{w}{3!}e_{ij}(\vec{k})
\beta^A_{ij}(\vec{k})\quad &
\bar{a}^p(\vec{k})\rightsquigarrow\frac{w}{3!}e_{ij}(\vec{k})
\bar{\beta}^A_{ij}(\vec{k})
\end{array}
\end{eqnarray*}
\begin{eqnarray*}
\begin{array}{ll}
\beta^{r+}_{ij}(\vec{k})\rightsquigarrow\beta^{\alpha}_{ij}(\vec{k})\quad
& \bar{\beta}^{r+}_{ij}(\vec{k})\rightsquigarrow
\bar{\beta}^{\alpha}_{ij}(\vec{k})\\
a^{r+}(\vec{k})\rightsquigarrow a^{\alpha}(\vec{k})\quad &
\bar{a}^{r+}(\vec{k})\rightsquigarrow\bar{a}^{\alpha}(\vec{k})\\
\sigma^{r+}_{i}(\vec{k})\rightsquigarrow
\sigma^{\alpha}_{i}(\vec{k})\quad &
\bar{\sigma}^{r+}_{i}(\vec{k})\rightsquigarrow
\bar{\sigma}^{\alpha}_{i}(\vec{k})\\
b^{r+}_{i}(\vec{k})\rightsquigarrow b^{\alpha}_{i}(\vec{k})\quad &
\bar{b}^{r+}_{i}(\vec{k})\rightsquigarrow \bar{b}^{\alpha}_{i}(\vec{k})\\
\alpha^{r+}_{i}(\vec{k})\rightsquigarrow
\alpha^{\alpha}_{i}(\vec{k})\quad &
\bar{\alpha}^{r+}_{i}(\vec{k})\rightsquigarrow
\bar{\alpha}^{\alpha}_{i}(\vec{k})
\end{array}
\end{eqnarray*}
\begin{eqnarray*}
\begin{array}{lll}
\beta^{r-}_{ij}(\vec{k})\rightsquigarrow
-i\frac{w}{3}e_{ij}(\vec{k})a^{\alpha}(\vec{k})&\quad &
\bar{\beta}^{r-}_{ij}(\vec{k})\rightsquigarrow
i\frac{w}{3}e_{ij}(\vec{k})\bar{a}^{\alpha}(\vec{k})\\
a^{r-}(\vec{k})\rightsquigarrow
-i\frac{w}{3!}e_{ij}(\vec{k})\beta^{\alpha}_{ij}(\vec{k})&\quad  &
\bar{a}^{r-}(\vec{k})\rightsquigarrow
i\frac{w}{3!}e_{ij}(\vec{k})\bar{\beta}^{\alpha}_{ij}(\vec{k})\\
\sigma^{r-}_{i}(\vec{k})\rightsquigarrow
\frac{w}{3}e_{ij}(\vec{k})\sigma^{\alpha}_j(\vec{k})& \quad &
\bar{\sigma}^{r-}_{i}(\vec{k})\rightsquigarrow
\frac{w}{3}e_{ij}(\vec{k})\bar{\sigma}^{\alpha}_j(\vec{k})\\
b^{r-}_i(\vec{k})\rightsquigarrow-\frac{w}{3}e_{ij}(\vec{k})[b^{\alpha}_j(\vec{k})+
\sigma^{\alpha}_j(\vec{k})]&\quad  &
\bar{b}^{r-}_i(\vec{k})\rightsquigarrow-
\frac{w}{3}e_{ij}(\vec{k})[\bar{b}^{\alpha}_j(\vec{k})+
\bar{\sigma}^{\alpha}_j(\vec{k})]\\
\alpha^{r-}_i(\vec{k})\rightsquigarrow
i\frac{w}{3}e_{ij}(\vec{k})[2b^{\alpha}_j(\vec{k})+
\sigma^{\alpha}_j(\vec{k})-i\alpha^{\alpha}_j(\vec{k})]& \quad &
\bar{\alpha}^{r-}_i(\vec{k})\rightsquigarrow-
i\frac{w}{3}e_{ij}(\vec{k})[2\bar{b}^{\alpha}_j(\vec{k})+
\bar{\sigma}^{\alpha}_j(\vec{k})+i\bar{\alpha}^{\alpha}_j(\vec{k})].
\end{array}
\end{eqnarray*}
In fact, by substituting the previous expressions into the
algebraic conditions and the symplectic structure for the $R=0$
case (given in appendix \ref{R=0}) we recover \emph{exactly}
equations (\ref{016}) and the symplectic structure given in the
previous section. This indicates that the energy-momentum density
can be also obtained from the one in $R=0$ by this procedure.
Doing so we find
\begin{eqnarray}
\tau^a\dd P_a=\dd
\int_{\mathbb{R}^3}\frac{d^3\vec{k}}{w}(\tau^ak_a)\bigg\{
16\overline{\beta}^A_{ij}(\vec{k}) \Re_{AB}\beta^B_{ij}(\vec{k})
+16 a^{\alpha}(\vec{k})\Re^-_{\alpha\beta}
\overline{a}^{\beta}(\vec{k})
-8\beta^{\alpha}_{ij}(\vec{k})\Re^+_{\alpha\beta}
\overline{\beta}^{\beta}_{ij}(\vec{k})& &\nonumber\\
-16\left[\beta^A_{ij}(\vec{k})\Im_{A\alpha}
\overline{\beta}^{\alpha}_{ij}(\vec{k})+\overline{\beta}^A_{ij}(\vec{k})
\Im_{A\alpha} \beta^{\alpha}_{ij}(\vec{k})\right]
-\frac{16}{3}we_{ij}(\vec{k})\left[\beta^A_{ij}(\vec{k})
\Re_{A\alpha}
\overline{a}^{\alpha}(\vec{k})-\overline{\beta}^A_{ij}(\vec{k})
\Re_{A\alpha} a^{\alpha}(\vec{k})\right]& &\nonumber\\
-\frac{8}{3}we_{ij}(\vec{k})\left[\beta^{\alpha}_{ij}(\vec{k})
\Im^-_{\alpha\beta}\overline{a}^{\beta}(\vec{k})
-\overline{\beta}^{\alpha}_{ij}(\vec{k}) \Im^-_{\alpha\beta}
a^{\beta}(\vec{k})\right]& &\nonumber\\
+16\left[[b^{\alpha}_i(\vec{k})-i\alpha^{\alpha}_i(\vec{k})]
\Im_{\alpha A}\overline{b}^A_i(\vec{k})
+[\overline{b}^{\alpha}_i(\vec{k})+i
\overline{\alpha}^{\alpha}_i(\vec{k})]
\Im_{\alpha A} b^A_i(\vec{k})\right]& & \nonumber\\
+\frac{8i}{3}we_{ij}(\vec{k})\left[\bar{\sigma}^a_i(\vec{k})
\Im^+_{\alpha\beta} b^{\beta}_j(\vec{k})+\sigma^a_i(\vec{k})]
\Im^+_{\alpha\beta} \bar{b}^{\beta}_j(\vec{k}) \right]
& & \nonumber\\
-8i\left[\alpha^{\alpha}_{i}(\vec{k}) \Re^+_{\alpha\beta}
\overline{b}^{\beta}_i(\vec{k})-\overline{\alpha}^{\alpha}_{i}(\vec{k})
\Re^+_{\alpha\beta}
b^{\beta}_i(\vec{k})\right]+16iwe_{ij}(\vec{k})
\bar{b}^{\alpha}_i(\vec{k})\Im^b_{\alpha\beta}
b^{\beta}_{j}(\vec{k})+16 \bar{b}^{\alpha}_i(\vec{k})
\Re^+_{\alpha\beta}b^{\beta}_i(\vec{k})\bigg\}& &\nonumber
\end{eqnarray}
\begin{eqnarray}
+(\tau^ak_a-\tau^0w)\bigg\{4\left[\bar{\sigma}^{\alpha}_i(\vec{k})
\Re^+_{\alpha\beta}
b^{\beta}_i(\vec{k})+\sigma^{\alpha}_i(\vec{k})
\Re^+_{\alpha\beta} \bar{b}^{\beta}_i(\vec{k})\right]
+8\left[\sigma^{\alpha}_i(\vec{k}) \Im_{\alpha A}
  \bar{b}^A_i(\vec{k})+\bar{\sigma}^{\alpha}_i(\vec{k})
\Im_{\alpha A} b^A_i(\vec{k})\right]& &\nonumber\\
-\frac{4}{3}iwe_{ij}(\vec{k})\left[\bar{\sigma}^{\alpha}_i(\vec{k})
\Im^-_{\alpha\beta}
b^{\beta}_j(\vec{k})+\sigma^{\alpha}_i(\vec{k})
\Im^-_{\alpha\beta} \bar{b}^{\beta}_j(\vec{k})\right]
+\frac{8}{3}iwe_{ij}(\vec{k})\bar{\sigma}^{\alpha}_i(\vec{k})
\Im^a_{\alpha\beta}\sigma^{\beta}_j(\vec{k})\bigg\}, & &
\nonumber
\end{eqnarray}
where $k^{\mu}=(w,\vec{k})$, and $k_0=-w$. The helicity of the
physical states can be obtained in a similar way by projecting the
angular momentum density along the direction of $\vec{k}$
\begin{eqnarray*}
\lambda(\vec{k})=
-8i\sigma^\alpha_i(\vec{k})\Im^b_{\alpha\beta}\bar{\sigma}^\beta_i(\vec{k})
-4i\sigma^\alpha_i(\vec{k})[\Im^+_{\alpha\beta}+2\Im_{\alpha\beta}^b]
\bar{b}^\beta_i(\vec{k})
+4i\bar{\sigma}^\alpha_i(\vec{k})[\Im^+_{\alpha\beta}+2\Im_{\alpha\beta}^b]
b^\beta_i(\vec{k})
&&\\
-\frac{4w}{3}e_{ij}(\vec{k})\sigma^\alpha_i(\vec{k})\Re^+_{\alpha\beta}
\bar{b}^\beta_j(\vec{k})+
\frac{4w}{3}e_{ij}(\vec{k})\bar{\sigma}^\alpha_i(\vec{k})\Re^+_{\alpha\beta}
b^\beta_j(\vec{k})-16ib^\alpha_i(\vec{k})
\Im^b_{\alpha\beta}\bar{b}^\beta_i(\vec{k})&&\\
-\frac{16w}{3}e_{ij}(\vec{k})b^\alpha_i(\vec{k})\Re^+_{\alpha\beta}\bar{b}^\beta_j(\vec{k})
+\frac{8iw}{3}e_{ij}(\vec{k})\alpha^\alpha_i(\vec{k})
\Re^+_{\alpha\beta}\bar{b}^\beta_j(\vec{k})
+\frac{8iw}{3}e_{ij}(\vec{k})\bar{\alpha}^\alpha_i(\vec{k})\Re^+_{\alpha\beta}
b^\beta_j(\vec{k})&&\\
-\frac{8w}{3}e_{ij}(\vec{k})[2b^\alpha_i(\vec{k})+\sigma^\alpha_i(\vec{k})-
2i\alpha^\alpha_i(\vec{k})]\Im_{\alpha A} \bar{b}^A_j(\vec{k})
+\frac{8w}{3}e_{ij}(\vec{k})[2\bar{
b}^\alpha_i(\vec{k})+\bar{\sigma}^\alpha_i(\vec{k})+
2i\bar{\alpha}^\alpha_i(\vec{k})]\Im_{\alpha A} b^A_j(\vec{k})&&\\
+8\alpha^\alpha_i(\vec{k})\Im^-_{\alpha\beta}
[\bar{b}^\beta_i(\vec{k})+\bar{\sigma}^\beta_i(\vec{k})]
+8\bar{\alpha}^\alpha_i(\vec{k})\Im^-_{\alpha\beta}
[b^\beta_i(\vec{k})+\sigma^\beta_i(\vec{k})].&&
\end{eqnarray*}

\section{\label{Semibound}Conditions to have a semibounded energy density}

The main physical requirement that we must impose to the kinetic
terms that we are considering in this paper is to have a
semibounded energy density. This is necessary in order to
guarantee stability once interactions are added in. In the absence
of algebraic conditions such as (\ref{016}) this would be
straightforward but here the situation is complicated by their
presence. Looking at the structure of the energy density we see
that some Fourier modes lack a quadratic part, that is, the matrix
representation of the energy as a quadratic form has some zero
entries in the main diagonal. This happens, for example, with the
fields $\alpha^{\alpha}_i(\vec{k})$ and
$\bar{\alpha}^{\alpha}_i(\vec{k})$. A well known fact in linear
algebra tells us that whenever this happens a quadratic form
cannot be either definite or semidefinite. This forces us to
impose suitable conditions on the $P$, $Q$, and $R$ matrices to
eliminate these terms (specifically the rows and columns of the
matrix where such diagonal zeroes appear). Here the presence of
the algebraic conditions (\ref{016}) forces us to study the
restriction of the energy to the vector subspace defined by them.
Before doing this it is convenient to rewrite the relevant part of
the energy in terms of the $+$ and $-$ objects introduced at the
end of appendix \ref{difforms} because, as also happened with the
algebraic conditions (\ref{017}), the $+$ and $-$ fields decouple.
It is also convenient to split the fields into their real and
imaginary parts to express the real energy in terms of real
quantities. The $\alpha$-dependent part of the energy density
reads
\begin{widetext}
\begin{eqnarray*}
\left[ \begin{array}{ll}\Re\alpha_-^{\alpha} &
\Im\alpha_-^{\alpha}\end{array} \right] \left[
\begin{array}{ccccrc} 0 & 4\Im^-_{\alpha\beta} & 4\Im^-_{\alpha\beta}
&-8\Im_{\alpha A} & -4\Re^+_{\alpha\beta} & 0 \\
8\Im_{\alpha A} & 4\Re^+_{\alpha\beta} & 0 & 0 &
4\Im^-_{\alpha\beta} & 4\Im^-_{\alpha\beta} \end{array}\right]
\left[  \begin{array}{c} \Re b_-^A \\ \Re b_-^{\beta} \\ \Re
\sigma_-^{\beta} \\  \Im b_-^A \\  \Im b_-^{\beta} \\
\Im\sigma_-^{\beta}\end{array}\right]\quad.
\end{eqnarray*}
\end{widetext}
The condition that this terms are absent on the vector subspace
defined by the algebraic conditions (\ref{0017}) is equivalent to
the condition that the rank of the following matrix coincides with
the rank of the submatrix defined by the the algebraic conditions
alone
\begin{widetext}
\begin{eqnarray*}
\left[
\begin{array}{crccrc} 0 & 4\Im^-_{\alpha\beta} & 4\Im^-_{\alpha\beta}
&-8\Im_{\alpha A} & -4\Re^+_{\alpha\beta} & 0 \\
8\Im_{\alpha A} & 4\Re^+_{\alpha\beta} & 0 & 0 &
4\Im^-_{\alpha\beta} & 4\Im^-_{\alpha\beta} \\
\hline -2 \Im_{\alpha A} & -2\Re^b_{\alpha\beta} &
\Re^-_{\alpha\beta} & -2\Re_{\alpha A} & 2\Im^b_{\alpha\beta} &
-\Im^-_{\alpha\beta} \\
2\Re^a_{\alpha\beta} & -2\Im^b_{\alpha\beta} & \Im^-_{\alpha\beta}
& -2\Im^a_{\alpha\beta} & -2\Re^b_{\alpha\beta} &
\Re^-_{\alpha\beta} \\
0 & \Re_{A\beta} & \Re_{A\beta} & -\Re_{AB} & \Im_{A\beta} & 0 \\
\Re_{AB} & -\Im_{A\beta} & 0 & 0 & \Re_{A\beta} & \Re_{A\beta}
\end{array}\right] \quad.
\end{eqnarray*}
\end{widetext}
Now it is straightforward to check that the full matrix is
non-singular; in fact, by adding the third column to the second,
the sixth to the fourth, and then the resulting first row to the
fourth, the second to the third, and finally rearranging the rows
and columns (multiplying, when necessary by the appropriate
non-zero factors) the resulting matrix is
\begin{eqnarray*}
\left[\begin{array}{cc} 0 & A \\ A & 0\end{array}\right]
\end{eqnarray*}
with
\begin{eqnarray*} A=
\left[\begin{array}{ccc} -\Re^+_{\alpha\beta}& \Im^-_{\alpha\beta} & 2\Im_{\alpha A} \\
\Im^+_{\alpha\beta} & \Re^-_{\alpha\beta} & 2\Re_{\alpha A}\\
\Im_{A\beta} & \Re_{A\beta} & \Re_{AB} \end{array}\right].
\end{eqnarray*}
This matrix is proportional to the matrix of $\mathcal{P}$ in the
vector subspace of $\mathbb{C}^{2N}$ spanned by the linearly
independent vectors
$$\left\{\left[\begin{array}{c} e_{\alpha} \\
\overline{e}_{\alpha} \end{array}\right],\,\,\left[\begin{array}{c} ie_{\alpha} \\
-i\overline{e}_{\alpha} \end{array}\right],\,\,\left[\begin{array}{c} iv_A \\
-i\overline{v}_A\end{array} \right]\right\}$$ and, hence, it is
non-singular. This means that the matrix that we started with is
non-singular and, hence, it is impossible that the first two rows
depend linearly on the remaining ones. The conclusion of this
analysis is that the only way to avoid having the $\alpha$-terms
is by working with matrices $P$, $Q$, and $R$ such that the
$e_{\alpha}$ sector is absent. This condition means that
$\textrm{rank}\mathcal{P}=N$, or in a more convenient form
$$\textrm{rank}\left[\begin{array}{cr} iR^t & Q \\
P & i R\end{array}\right]=N.$$ Notice that in the $R=0$ case this
is just $\textrm{rank} P+\textrm{rank} Q=N$ or $\dim\ker
P+\dim\ker Q=N$. Once we impose this condition we get that the
symplectic structure is simply:
\begin{eqnarray*}
\Omega=16i\int_{\mathbb{R}^3}\frac{d^3k}{w}
\dd\overline{\beta}^A_{ij}(\vec{k})
\Re_{AB}\ww\dd\beta^B_{ij}(\vec{k})& &\nonumber
\end{eqnarray*}
and the energy-momentum density becomes
\begin{eqnarray*}
\tau^a\dd P_a=\dd
\int_{\mathbb{R}^3}\frac{d^3\vec{k}}{w}(\tau^ak_a)
\,16\overline{\beta}^A_{ij}(\vec{k})
\Re_{AB}\beta^B_{ij}(\vec{k}).& &\nonumber
\end{eqnarray*}
The helicity density is identically zero, i.e. in the physically
consistent case the action (\ref{002}) describes only scalar
particles. In order to ensure that the energy density is positive
definite we must require that the non-singular matrix $\Re_{AB}$
be definite\footnote{If $\Re_{AB}=v_-^{At}\mathcal{P}v_-^B$ were
singular we should be able to find $\lambda_A \in\mathbb{R}$ such
that $v_-^{At}\mathcal{P}(v_-^B\lambda_B)=0$; as the $v_-^A$ are
linearly independent this means that $(v_-^B\lambda_B)\in \ker
\mathcal{P}$. As it obviously belongs to $ \ker \mathcal{Q}$, and
$ \ker \mathcal{P}\cap\ker \mathcal{Q}=\{0\}$, $(v_-^B\lambda_B)$
must be zero and also $\lambda_A$.} Notice, also, that the
symplectic structure is non-degenerate ($\Re_{AB}$ is always
non-singular) and that it is possible to simultaneously
diagonalize both the symplectic structure and the energy density.
Knowing that the $e_{\alpha}$ sector must be absent we can use an
internal basis of $\mathbb{C}^{2N}$ spanned
by the vectors $\left[\begin{array}{c} v_A \\
\overline{v}_A\end{array} \right]$ and $\left[\begin{array}{c} v_A \\
-\overline{v}_A\end{array} \right]$ and expand
$$
\mathcal{B}=*B^A\left[\begin{array}{c} v_A \\
\overline{v}_A\end{array} \right]-iB^A\left[\begin{array}{c} v_A \\
-\overline{v}_A\end{array} \right] $$ that introduced back in
(\ref{018}) gives
$$
S_2^{\prime}=-2\int_{\mathbb{R}^4}dB^{A} \wedge*\Re_{AB}dB^B.$$

Now as $\Re_{AB}$ is actually a positive definite symmetric we
know that we can diagonalize it to be the identity and, hence, the
previous action is a sum of N Maxwell actions as in the $R=0$
case.

\section{\label{examples} The $R=0$ case}

In this section we check that we recover the previously known
results for the $R=0$ case within the formalism presented in this
paper. The first step is to compute $v_A$ by finding the kernel of
$$\mathcal{P}=\left[\begin{array}{cc} P+Q & P-Q \\P-Q & P+Q
\end{array} \right]$$
It is straightforward to see that the elements of
$\ker\mathcal{P}$ must have the form
$$\lambda_p\left[\begin{array}{c} e_p \\ e_p\end{array}\right]+
\lambda_q\left[\begin{array}{c} e_q \\ -e_q\end{array}\right].$$
where $e_p\in \ker P$,  $e_q\in \ker Q$ are \emph{real} vectors
 \footnote{Here we
follow the notation used in appendix \ref{R=0}} and
$\lambda_p,\,\lambda_Q\in\mathbb{R}$. As we are taking the vectors
in $\ker\mathcal{P}$ and
$\ker\mathcal{Q}$ in the form  $\left[\begin{array}{c} v_A \\
\overline{v}_A\end{array} \right]$ and $\left[\begin{array}{c} v_A \\
-\overline{v}_A\end{array} \right]$ we must use the following
linearly independent vectors\footnote{the $1/2$ factor is included
to take into account the different choices of constants in
(\ref{001}) and (\ref{002})}
$$\left\{\frac{1}{2}\left[\begin{array}{c} e_p \\ e_p\end{array} \right],\,\,
\frac{1}{2}\left[\begin{array}{c} ie_q \\ -ie_q\end{array}
\right]\right\}$$ which, in practice means that we have two types
of $v_A$ objects: the real $e_p$ and the purely imaginary $ie_q$.
We complete this set with
$$\left\{\frac{1}{2}\left[\begin{array}{c} e_r \\ e_r\end{array} \right]\right\}$$
where the $e_r$ are also chosen to be real and together with
$e_p$, $e_q$ give a basis of $\mathbb{R}^N$. In the following we
will use indices $p$, $q$ instead of $A$, and $r$ instead of
$\alpha$.

 We compute now the several internal matrices appearing in
the expressions of the symplectic structure, the energy-momentum
density and the helicity density
\begin{eqnarray*}
& \Re_{AB}\equiv \frac{1}{4}\left[\begin{array}{cc} e_p^tQe_p & 0 \\
0 & -e_q^tPe_q\end{array}\right],\quad & \nonumber\\
& \Im_{A\alpha}\equiv \frac{1}{4}e_q^tPe_r, \quad \Im_{\alpha A
}\equiv \frac{1}{4}e_r^tPe_q,\quad \Re_{A\alpha}\equiv
\frac{1}{4}e_p^tQe_r ,\quad \Re_{\alpha A }\equiv
\frac{1}{4}e_r^tQe_p,\quad & \nonumber\\
& \Re^+_{\alpha\beta}\equiv \frac{1}{2}e_r^tPe_r,\quad
\Re^-_{\alpha\beta}\equiv \frac{1}{2}e_r^tQe_r,\quad
\Im^+_{\alpha\beta}\equiv0\quad \Im^-_{\alpha\beta}\equiv0,\quad
\Im^b_{\alpha\beta}\equiv0,\quad \Im^a_{\alpha\beta}\equiv0,\quad
& \nonumber
\end{eqnarray*}
we get then
\begin{eqnarray*}
\Omega=4\!\int_{\mathbb{R}^3}\frac{d^3\vec{k}}{w}
\left\{i\dd\bar{\beta}^p_{ij}(\vec{k}) e_p^tQe_{p^{\prime}}\ww\dd
\beta^{q^{\prime}}_{ij}(\vec{k}) -i\dd\bar{\beta}^q_{ij}(\vec{k})
e_q^tPe_{q^{\prime}}\ww\dd \beta^{q^{\prime}}_{ij}(\vec{k})
+i\dd\beta^q_{ij}(\vec{k})e_qPe_r\ww\dd\bar{\beta}^r_{ij}(\vec{k})\right.
& & \nonumber\\
-i\dd\bar{\beta}^q_{ij}(\vec{k})e_q^tPe_r\ww\dd\beta^r_{ij}(\vec{k})
+i\dd\beta^r_{ij}(\vec{k})e_r^{t}Pe_{r}^{\prime}\ww\dd
\bar{\beta}^{r^{\prime}}_{ij}(\vec{k}) -2i\dd
a^r(\vec{k})e_r^tQe_{r^{\prime}}\ww\dd\bar{a}^{r^{\prime}}(\vec{k})
& & \nonumber\\
+\frac{1}{3}iwe_{ij}(\vec{k})\left[\dd\beta^p_{ij}(\vec{k})
e_p^tQe_r\ww\dd\bar{a}^r(\vec{k})+\dd\bar{\beta}^p_{ij}(\vec{k})
e_p^tQe_r\ww\dd a^r(\vec{k})\right]
& & \nonumber\\
-\frac{1}{2}i\left[\dd\sigma^r_i(\vec{k})e^tPe_{r^{\prime}}\ww\dd
\bar{b}^{r^{\prime}}_i(\vec{k})-\dd\bar{\sigma}^r_i(\vec{k})
e^tPe_{r^{\prime}}\ww\dd b^{r^{\prime}}_i(\vec{k}) \right]
& & \nonumber\\
-\frac{1}{2}\left[2\dd b^r_i(\vec{k})+\dd
\sigma^r_i(\vec{k})-2i\dd\alpha^r_i(\vec{k})
\right]e_r^tPe_q\ww\dd\bar{b}^q_i(\vec{k})& &
\nonumber\\
+\frac{1}{2}\left[2\dd \bar{b}^r_i(\vec{k})+\dd
\bar{\sigma}^r_i(\vec{k})+2i\dd\bar{\alpha}^r_i(\vec{k})
\right]e_r^tPe_q\ww\dd b^q_i(\vec{k})& & \nonumber\\
\left. -\dd \bar{\alpha}^r_i(\vec{k})e_r^tPe_{r^{\prime}}\ww\dd b
^{r^{\prime}}_i(\vec{k})-\dd
\alpha^r_i(\vec{k})e_r^tPe_{r^{\prime}}
\ww\dd\bar{b}^{r^{\prime}}_i(\vec{k})-2i\dd
b^r_i(\vec{k})e_r^tPe_{r^{\prime}}
\ww\dd\bar{b}^{r^{\prime}}_i(\vec{k})\right\}. & & \nonumber
\end{eqnarray*}
For the energy-momentum we find
\begin{eqnarray}
\tau^a\dd P_a=\dd
\int_{\mathbb{R}^3}\frac{d^3\vec{k}}{w}(\tau^ak_a)\bigg\{4\bar{\beta}^p_{ij}(\vec{k})
e_p^tQe_{p^{\prime}}\beta^{p^{\prime}}_{ij}(\vec{k})-4\bar{\beta}^q_{ij}(\vec{k})
e_q^tPe_{q^{\prime}}\beta^{q^{\prime}}_{ij}(\vec{k})-4\bar{\beta}^r_{ij}(\vec{k})
e_r^tPe_{r^{\prime}}\beta^{r^{\prime}}_{ij}(\vec{k})-&
&\nonumber\\
-4[\bar{\beta}^q_{ij}(\vec{k})
e_q^tPe_r\beta^r_{ij}(\vec{k})+\beta^q_{ij}(\vec{k})
e_q^tPe_r\bar{\beta}^r_{ij}(\vec{k})] & &\nonumber\\
-\frac{4}{3}we_{ij}(\vec{k})[
\beta^p_{ij}(\vec{k})e_p^tQe_r\bar{a}(\vec{k})-
\bar{\beta}^p_{ij}(\vec{k})e_p^tQe_ra(\vec{k})] &
&\nonumber\\
+4\left\{[b^r_i(\vec{k})-i\alpha^r_i(\vec{k})]
e_r^tPe_q\bar{b}^q_i(\vec{k})+
[\bar{b}^r_i(\vec{k})+i\bar{\alpha}^r_i(\vec{k})]
e_r^tPe_qb^q_i(\vec{k})\right\} & &\nonumber\\+8a^r(\vec{k})e_r^tQ
e_{r^{\prime}}\bar{a}^{r^{\prime}}_i(\vec{k})+8\bar{b}^r_i(\vec{k})e_r^t
Pe_{r^{\prime}}b^{r^{\prime}}_i(\vec{k})
& &\nonumber\\
-4i[\alpha^r_i(\vec{k})e_r^t
Pe_{r^{\prime}}\bar{b}^{r^{\prime}}_i(\vec{k})-
\bar{\alpha}^r_i(\vec{k})e_r^t
Pe_{r^{\prime}}b^{r^{\prime}}_i(\vec{k})]
& & \nonumber\\
+2(\tau^ak_a-\tau^0w) \left[\sigma^r_i(\vec{k})e_r^tP
[e_q\bar{b}^q(\vec{k})+e_{r^{\prime}}\bar{b}^{r^{\prime}}(\vec{k})]+
\bar{\sigma}^r_i(\vec{k})e_r^tP
[e_qb^q(\vec{k})+e_{r^{\prime}}b^{r^{\prime}}(\vec{k})]\right]\bigg\}&
& \nonumber
\end{eqnarray}
and the helicity density is
\begin{eqnarray}
\lambda(\vec{k})=\epsilon_{ijk}\frac{k^i}{w}\left\{
-8i\bar{b}^r_j(\vec{k})e_r^{t}Pe_{r^{\prime}}b_k^{r^{\prime}}(\vec{k})
-4i\bar{b}^r_j(\vec{k})e_r^{t}Pe_qb_k^q(\vec{k})
-4i\bar{b}^q_j(\vec{k})e_q^{t}Pe_rb_k^r(\vec{k}) \right.& &
\nonumber\\
\left.-2i[\bar{\sigma}^r_j(\vec{k})+2i\bar{\alpha}^r_j(\vec{k})]Pb_k(\vec{k})+
2i[\sigma^r_j(\vec{k})-2i\alpha^r_j(\vec{k})]P\bar{b}_k(\vec{k})\right\}.&
&  \nonumber
\end{eqnarray}

All these expressions coincide with the ones given in appendix
\ref{R=0} after making the identifications
\begin{eqnarray*}
\alpha^p_i(\vec{k})\equiv\frac{2}{3}we_{ij}(\vec{k})b^p_j(\vec{k})
& , &
\beta^p_{ij}(\vec{k})\equiv\frac{1}{3}we_{ij}(\vec{k})a^p(\vec{k})
\end{eqnarray*}
and using the algebraic conditions (\ref{016}).
\section{\label{conclusions}Conclusions}

The main purpose of the paper is to classify all the possible
kinetic terms for 2-form fields in four dimensions completing
previous work in this direction. This classification may be useful
to understand quantum field theories involving the coupling of
1-forms and 2-forms. From the technical point of view several
issues should be emphasized. First, in order to disentangle the
algebraic structure of the space of solutions to the field
equations (and have the possibility of using previous results to
solve the necessary conditions that appear as an intermediate step
in the process of solving them) we have introduced a set of
\emph{complex} projection operators.  Their use allow us to
circumvent the difficulties associated with the fact that the
Hodge dual and the exterior derivative (and its dual) do not
commute. Second, the fact that we have introduced complex objects
forced us to impose suitable reality conditions to describe real
solutions to the field equations. This has a intriguing similarity
with the original Ashtekar \cite{Ashtekar:1987gu, Ashtekar:1986yd}
formulation of General Relativity as derived from the self-dual
action \cite{Jacobson:1988yy, Samuel:1987yy} where reality
conditions were a key ingredient to keep the structure of the
Hamiltonian constraint as simple as possible.

The main physical requirement that we impose on the models
described by the action (\ref{002}) is that the energy must be
semibounded. This is necessary to guarantee stability once
interactions are included.  This simple condition strongly
constrains the physical content of these actions: as we have shown
all of them can be rewritten by means of linear, non-singular
field redefinitions as the sum of several Maxwell 2-form actions
describing massless scalars. This result is important, among other
things, to study the possible deformations of general 2-form
theories by using the techniques developed by Henneaux and
collaborators \cite{Henn, HennK}. We want to stress that the main
difficulty that we have to overcome is to find \textit{sufficient}
and \textit{necessary} conditions on $P$, $Q$, and $R$ that
guarantee the semiboundedness of the energy. Although it is easy
to give sufficient conditions it is much harder to find those that
are also necessary; this is the main problem solved in the paper.

The approach that we follow gives a partial classification of
quantum field theories --at least of those that are ``continuosly"
connected to their kinetic terms-- which are also, in a sense,
those that can be expected to allow a consistent perturbative
treatment (around the zero value of the fields). It is important
to realize that gauge symmetries appear in a natural way as given
by those functions appearing in the solution to the field
equations that do not appear in the simplectic structure. Their
presence or absence in the kinetic terms that we consider is
dictated only by the conditions imposed on the energy; and their
extension to a full interacting theory will be given by their
consistent deformations.

The next step in the process of classifying kinetic terms consists
of considering quadratic terms involving different types of
differential forms, such as 1-forms and 2-forms. The techniques
developed in this and previous papers will certainly be of much
help to disentangle the presumably complicated algebraic structure
of these more elaborate models.

\appendix
\section{\label{difforms}Conventions for differential forms}

We give in this appendix our definitions and conventions for
differential forms in four dimensions. A $s$-form $\omega$ defined
on a $N$ dimensional differentiable manifold $\mathcal{M}$
(endowed with coordinates $x^a$) is given by
\begin{eqnarray*}
\omega(x)=\omega_{a_1\cdots a_s}(x)dx^{a_1}\wedge\cdots\wedge
dx^{a_s}
\end{eqnarray*}
with
\begin{eqnarray*}
\omega_{a_1\cdots a_s}=\omega_{[a_1\cdots a_s]}\equiv \frac{1}{s!}
\sum_{\pi\in{\cal S}_s}(-1)^\pi\omega_{\pi(a_1)\cdots \pi(a_s)}
\nonumber
\end{eqnarray*}
and $\pi\in\mathcal{S}_s$ a permutation of order s. The space of
$s$-forms on $\mathcal{M}$ will be denoted as
$\Omega_s(\mathcal{M})$.

The exterior (wedge) product of a $s$-form $\omega$ and a $r$-form
$\xi$ is defined by
\begin{eqnarray*}
\omega\wedge \xi =\omega_{[a_1\cdots a_s}\xi_{b_1\cdots
b_r]}dx^{a_1}\wedge\cdots\wedge dx^{a_s}\wedge
dx^{b_1}\wedge\cdots \wedge dx^{b_r}
\end{eqnarray*}
and satisfies
\begin{eqnarray*}
\omega\wedge \xi&=&(-1)^{sr}\xi\wedge \omega,\\
(\xi\wedge\eta)\wedge \omega&=& \xi\wedge (\eta\wedge \omega).
\end{eqnarray*}
The exterior differential that takes a $s$-form $\omega$ to a
$(s+1)$-form is defined as
\begin{eqnarray*}
d\omega=\partial_{[a_1}\omega_{a_2\cdots
a_{s+1}]}dx^{a_1}\wedge\cdots \wedge dx^{a_{s+1}},
\end{eqnarray*}
and has the following properties
\begin{eqnarray*}
d^2&=&0\\
d(\omega\wedge\xi)&=& d \omega\wedge\xi+(-1)^s\omega\wedge d\xi.
\end{eqnarray*}

In the presence of a non-degenerate metric in $\cal M$ we can
define the Hodge dual of a $s$-form $\omega$ as the $(N-s)$-form
given by
\begin{widetext}
\begin{eqnarray*}
*\omega=\frac{1}{(N-s)!}\frac{1}{\sqrt{|\det g|}}\omega_{b_1\cdots
b_s}\tilde{\eta}^{b_1\cdots b_sc_1\cdots c_{N-s}}g_{a_1c_1}\cdots
g_{a_{N-s}c_{N-s}}dx^{a_1}\wedge\cdots\wedge dx^{a_{N-s}}\, ,
\end{eqnarray*}
\end{widetext}
where $\tilde{\eta}^{b_1\cdots b_N}$ is the Levi-Civita tensor
density on $\cal M$ defined  to  be $+1$ for even permutations of
the indices and $-1$ for odd permutations, in any coordinate
chart. If $g_{ab}$ has Riemannian signature we have
\begin{eqnarray*}
**\omega=(-1)^{s(N-s)}\omega
\end{eqnarray*}
whereas for Lorentzian signatures we get
\begin{eqnarray*}
**\omega=(-1)^{s(N-s)+1} \omega\quad.
\end{eqnarray*}
In the presence of a metric it is also possible to define the
adjoint exterior differential $\delta$
\begin{eqnarray*}
\delta &=&(-1)^{N(s+1)+1}*d*\quad {\rm Riemannian \,\, signature}\\
\delta &=& (-1)^{N(s+1)}*d*\quad \quad{\rm Lorentzian \,\,
signature}.
\end{eqnarray*}
It takes $s$-forms to $(s-1)$-forms according to
\begin{eqnarray*}
\delta\omega=-s\nabla^a\omega_{aa_1\cdots a_{s-1}}d x^{a_1}\wedge
\cdots \wedge d x^{a_{s-1}}\quad,
\end{eqnarray*}
where $\nabla$ is the metric compatible, torsion-free, covariant
derivative and
 satisfies
\begin{eqnarray*}
\delta^2=0\quad.
\end{eqnarray*}
We introduce now the operator (Laplacian for Riemannian signatures
and wave operator for Lorentzian signatures)
\begin{eqnarray*}
\square= d\delta +\delta d\quad,
\end{eqnarray*}
that takes $s$-forms to $s$-forms and is given by
\begin{eqnarray*}
\square \omega=-\nabla_a\nabla^a\omega_{a_1\cdots
a_s}dx^{a_1}\wedge\cdots \wedge d x^{a_s};
\end{eqnarray*}
it commutes with both $d$ and $\delta$. Finally, let $X$ be a
vector field given by $X=X^a\partial_a$, we define the interior
product of $X$ and a $s$-form $\omega$ as the $(s-1)$-form
\begin{eqnarray*}
i_X\omega=sX^a\omega_{a a_2,\ldots,
a_{s-1}}dx^{a_2}\wedge\ldots\wedge dx^{a_{s-1}};
\end{eqnarray*}
it satisfies
\begin{eqnarray*}
(di_X+i_Xd)\omega=\mathcal{L}_X\omega;
\end{eqnarray*}
where $\mathcal{L}_X\omega$ is the Lie derivative of $\omega$
defined by the vector field $X$.

We will write the components of differential forms in a
Minkowskian background as
\begin{eqnarray*}
\omega=\left\{\begin{array}{c}\omega_{i_1\cdots
i_s}\\
\omega_{0i_1\cdots i_{s-1}}
\end{array}\right\}.
\end{eqnarray*}
The Fourier transforms for $s$-forms $\omega_s$ can be
parametrized as follows
\begin{eqnarray*}
\omega_0=
\left\{\begin{array}{c}\beta(\vec{k},t)\\ \\
0
\end{array}\right\}
\end{eqnarray*}
\begin{eqnarray*}
\omega_1=
\left\{\begin{array}{c}ik_i\alpha(\vec{k},t)+\beta_i(\vec{k},t)\\ \\
b(\vec{k},t)
\end{array}\right\}
\end{eqnarray*}
\begin{eqnarray*}
\omega_2=
\left\{\begin{array}{c}ik_{[i}\alpha_{j]}(\vec{k},t)+\beta_{ij}(\vec{k},t)\\ \\
ik_i a(\vec{k},t)+b_i(\vec{k},t)
\end{array}\right\}
\end{eqnarray*}
\begin{eqnarray*}
\omega_3=
\left\{\begin{array}{c}ik_{[i}\alpha_{jk]}(\vec{k},t)\\ \\
ik_{[i} a_{j]}(\vec{k},t)+b_{ij}(\vec{k},t)
\end{array}\right\}
\end{eqnarray*}
\begin{eqnarray*}
\omega_4=
\left\{\begin{array}{c}0\\ \\
ik_{[i} a_{jk]}(\vec{k},t)
\end{array}\right\}
\end{eqnarray*}
where $\alpha_{ij}$, $\beta_{ij}$, and $b_{ij}$ are antisymmetric
and transverse, and $a_i$, $b_i$, $\alpha_i$, and $\beta_i$ are
transverse.

We need also the expressions for the Hodge duals of these objects.
In the following it is useful to introduce the transverse
antisymmetric object $e_{ij}(\vec{k})$ defined by
\begin{eqnarray*}
e_{ij}(\vec{k})\equiv-\frac{3i}{w^2}\varepsilon_{ijk} k^k
\end{eqnarray*}
that satisfies the following properties
\begin{eqnarray}
& & \varepsilon_{ijk}= ik_{[i}e_{jk]}(\vec{k})\nonumber\\
& & e_{ij}(\vec{k})=\overline{e}_{ij}(-\vec{k})\nonumber\\
& & e_{ij}(\vec{k})
e_{jk}(\vec{k})=\frac{9}{w^2}\left(\delta_{ik}-\frac{k_i
k_k}{w^2}\right)\nonumber\\
& & e_{ij}(\vec{k}) e_{ij}(-\vec{k})=\frac{18}{w^2}\nonumber
\end{eqnarray}
In Fourier transform we have
\begin{eqnarray*}
*\omega_4=
\left\{\begin{array}{c}-\frac{4}{3}w^2e_{ij}(\vec{k})a_{ij}(\vec{k},t)\\ \\
0
\end{array}\right\}
\end{eqnarray*}
\begin{eqnarray*}
*\omega_3=
\left\{\begin{array}{c}ik_ie_{jk}(\vec{k})b_{jk}(\vec{k},t)+
w^2e_{ij}(\vec{k})a_j(\vec{k},t)\\ \\
-\frac{w^2}{3}e_{ij}(\vec{k})\alpha_{ij}(\vec{k},t)
\end{array}\right\}
\end{eqnarray*}
\begin{eqnarray*}
*\omega_2=
\left\{\begin{array}{c}\frac{2i}{3}k_{[i}e_{j]k}(\vec{k})b_k(\vec{k},t)-
\frac{w^2}{3}e_{ij}(\vec{k})a(\vec{k},t)\\ \\
-\frac{i}{3!}k_ie_{jk}(\vec{k})\beta_{jk}(\vec{k},t)-
\frac{w^2}{3!}e_{ij}(\vec{k})\alpha_j(\vec{k},t)
\end{array}\right\}
\end{eqnarray*}
\begin{eqnarray*}
*\omega_1=
\left\{\begin{array}{c}\frac{i}{6}k_{[i}e_{jk]}(\vec{k})b(\vec{k},t)
\\ \\
\frac{2i}{3\cdot3!}k_{[i}e_{j]k}(\vec{k})\beta_{k}(\vec{k},t)-
\frac{w^2}{3\cdot3!}e_{ij}(\vec{k})\alpha(\vec{k},t)
\end{array}\right\}
\end{eqnarray*}
\begin{eqnarray*}
*\omega_0=\left\{\begin{array}{c}
 0 \\ \\
-\frac{i}{4!}k_{[i}e_{jk]}(\vec{k})\beta(\vec{k},t)
\end{array}\right\}\quad.
\end{eqnarray*}
The objects appearing in solutions to the necessary conditions
(\ref{012}) can be parametrized \cite{edufersform} as
\begin{eqnarray*}
\gamma^A=\left\{\begin{array}{c}\beta^A_{ij}(\vec{k})e^{-iwt}+
\bar{\beta}^A_{ij}(-\vec{k})e^{iwt}\\
\\b^A_{i}(\vec{k})e^{-iwt}+\bar{b}^A_{i}(-\vec{k})e^{iwt}\end{array}\right\}.
\end{eqnarray*}
\begin{eqnarray*}
\gamma^{\alpha}=
\left\{\begin{array}{c}\beta^{\alpha}_{ij}(\vec{k})e^{-iwt}+
\bar{\beta}^{\alpha}_{ij}(-\vec{k})e^{iwt}\\ \\
\frac{i}{w}k_i\left[a^{\alpha}(\vec{k})e^{-iwt}+\bar{a}^{\alpha}
(-\vec{k})e^{iwt}\right]+b_i^{\alpha}(\vec{k})e^{-iwt}+\bar{b}^{\alpha}_i(-\vec{k})e^{iwt}
\end{array}\right\}
\end{eqnarray*}
\begin{eqnarray*}
d\delta\gamma_{\scriptscriptstyle {H}}&=&\left\{\begin{array}{l}
   \quad\quad\frac{i}{w}k_{[i}
    \left[\alpha_{j]}^{\alpha}(\vec{k})+wt\sigma_{j]}^{\alpha}
    (\vec{k})\right]e^{-iwt}+\frac{i}{w}k_{[i}\left[\bar{\alpha}_{j]}^{\alpha}(-\vec{k})+wt\bar{\sigma}_{j]}^{\alpha}
 (-\vec{k})\right]e^{iwt}\\  \\
\frac{1}{2}\left[
    -i\alpha_{i}^{\alpha}(\vec{k})+(1-iwt)\sigma_{i}^{\alpha}(\vec{k})\right]e^{-iwt}
    +\frac{1}{2}\left[i\bar{\alpha}_{i}(-\vec{k})+(1+iwt)\bar{\sigma}_{i}^{\alpha}(-\vec{k})\right]e^{iwt}
\end{array}\right\}
\end{eqnarray*}
We will also need
\begin{eqnarray*}
d\gamma^A=\left\{\begin{array}{c}ik_{[i}\beta^A_{jk]}(\vec{k})e^{-iwt}+
ik_{[i}\overline{\beta}^A_{jk]}(-\vec{k})e^{iwt}\\
\\-\frac{2}{3}ik_{[i}\left[b^A_{j]}(\vec{k})e^{-iwt}+
\overline{b}^A_{j]}(-\vec{k})e^{iwt}\right]-\frac{w}{3}i
\left[\beta_{ij}^A(\vec{k})e^{-iwt}-
\overline{\beta}_{ij}^A(-\vec{k})e^{iwt}\right]
\end{array}\right\}
\end{eqnarray*}
\begin{eqnarray*}
d\gamma^{\alpha}=\left\{\begin{array}{c}ik_{[i}
\beta^{\alpha}_{jk]}(\vec{k})e^{-iwt}+
ik_{[i}\overline{\beta}^{\alpha}_{jk]}(-\vec{k})e^{iwt}\\
\\-\frac{2}{3}ik_{[i}\left[b^{\alpha}_{j]}(\vec{k})e^{-iwt}+
\overline{b}^{\alpha}_{j]}(-\vec{k})e^{iwt}\right]-
\frac{w}{3}i\left[\beta_{ij}^{\alpha}(\vec{k})e^{-iwt}-
\overline{\beta}_{ij}^{\alpha}(-\vec{k})e^{iwt}\right]
\end{array}\right\}
\end{eqnarray*}
\begin{eqnarray*}
*\gamma^A=\left\{\begin{array}{c}
\frac{2}{3}ik_{[i}e_{j]k}(\vec{k})\left[b^A_k(\vec{k})e^{-iwt}+
\overline{b}^A_k(-\vec{k})e^{iwt}\right]
\\
\\-\frac{i}{3!}k_{i}e_{jk}(\vec{k})\left[\beta^A_{jk}(\vec{k})e^{-iwt}+
\overline{\beta}^A_{ij}(-\vec{k})e^{iwt}\right]
\end{array}\right\}
\end{eqnarray*}
\begin{eqnarray*}
*\gamma^{\alpha}=\left\{\begin{array}{c}
\frac{2}{3}ik_{[i}e_{j]k}(\vec{k})\left[b^{\alpha}_k(\vec{k})e^{-iwt}+
\overline{b}^{\alpha}_k(-\vec{k})e^{iwt}\right]-\frac{w}{3}e_{ij}(\vec{k})\left[a^{\alpha}(\vec{k})e^{-iwt}+
\overline{a}^{\alpha}(-\vec{k})e^{iwt}\right]
\\
\\-\frac{i}{3!}k_{i}e_{jk}(\vec{k})\left[\beta^{\alpha}_{jk}(\vec{k})e^{-iwt}+
\overline{\beta}^{\alpha}_{ij}(-\vec{k})e^{iwt}\right]
\end{array}\right\}
\end{eqnarray*}
\begin{eqnarray*}
*d\delta\gamma^{\alpha}_{\scriptscriptstyle{H}}=\left\{
\begin{array}{c} \frac{i}{3}k_{[i}e_{j]k}(\vec{k})
\left[\{(1-iwt)\sigma^{\alpha}_k(\vec{k})-i\alpha_k^{\alpha}(\vec{k})\}e^{-iwt}+
\{(1+iwt)\overline{\sigma}^{\alpha}_k(-\vec{k})+
i\overline{\alpha}_k^{\alpha}(-\vec{k})\}e^{iwt}\right]
\\ \\-\frac{w}{3!}e_{ij}(\vec{k})
\left\{[\alpha^{\alpha}_{j}(\vec{k})+wt\sigma^{\alpha}_j(\vec{k})]e^{-iwt}+[\overline{\alpha}^{\alpha}_{j}(-\vec{k})+
wt\overline{\sigma}^{\alpha}_j(-\vec{k})]e^{iwt}\right\}
\end{array}\right\}
\end{eqnarray*}
\begin{eqnarray*}
d\!*\!\gamma^{\alpha}=\left\{\begin{array}{c}-\frac{w}{3}ik_{[i}
e_{jk]}(\vec{k})
[a^{\alpha}(\vec{k})e^{-iwt}+\overline{a}^{\alpha}(-\vec{k})e^{iwt}]\\
\\\frac{2w}{9}k_{[i}e_{j]k}(\vec{k})[-ib^{\alpha}_k(\vec{k})e^{-iwt}+
i \overline{b}^{\alpha}_k(-\vec{k})e^{iwt}]
+\frac{iw^2}{9}e_{ij}(\vec{k})[a^{\alpha}(\vec{k})e^{-iwt}-
\overline{a}^{\alpha}(-\vec{k})e^{iwt}]
\end{array}\right\}
\end{eqnarray*}
\begin{eqnarray*}
*\delta
d\delta\gamma^{\alpha}_{\scriptscriptstyle{H}}=\left\{\!\begin{array}{c}0
\\
\\\frac{2}{9}iwk_{[i}e_{j]k}(\vec{k})[-i\sigma^{\alpha}_k(\vec{k})e^{-iwt}+
i\overline{\sigma}(-\vec{k})e^{iwt}]
\end{array}\!\right\}\quad.
\end{eqnarray*}

The expressions for the algebraic conditions derived from the
field equations, the symplectic form, the energy momentum density,
and the helicity density can be simplified by introducing a pair
of \emph{real}, $\vec {k}$-dependent, transverse vectors
$e^{\lambda}_i(\vec {k})$ ($\lambda=L,\,R$) satisfying the
following properties:
\begin{eqnarray*}
 & & k_ie^{\lambda}_i(\vec {k})=0,\\
 & & \{\vec{k},\vec{e}^R(\vec{k}),\vec{e}^L(\vec{k})\}\,\,\textrm{positive\,\,oriented},\\
 & & e^{\lambda}_i(\vec {k})e^{\mu}_i(\vec
 {k})=\delta^{\lambda\mu}\\
 & & e_{ij}(\vec {k})e^{\lambda}_j(\vec
 {k})=-\frac{3i}{\omega}\epsilon^{\lambda}_{\,\,\mu}e^{\mu}_i(\vec
 {k})\quad,
 \end{eqnarray*}
where $\epsilon^{\lambda}_{\,\,\mu}$ is the antisymmetric matrix
\begin{eqnarray*}
\left[
   \begin{array}{rr}
    \,\, 0 & \quad 1 \\
    \,\, -1 & \quad 0
   \end{array}\right]\quad,
\end{eqnarray*}
and writing
\begin{eqnarray*}
& & \beta^A_{ij}(\vec{k})=i\epsilon_{ijk}\frac{k_k}{\omega}\beta^A(\vec{k})\\
& & \beta^{\alpha}_{ij}(\vec{k})=i\epsilon_{ijk}\frac{k_k}{\omega}\beta^{\alpha}(\vec{k})\\
& & b^A_i(\vec{k})=e^{\lambda}_i(\vec{k})b^A_{\lambda}(\vec{k})\\
& & b^{\alpha}_i(\vec{k})=e^{\lambda}_i(\vec{k})b^{\alpha}_{\lambda}(\vec{k})\\
& & \alpha^{\alpha}_i(\vec{k})=e^{\lambda}_i(\vec{k})\alpha^{\alpha}_{\lambda}(\vec{k})\\
& & \sigma^{\alpha}_i(\vec{k})=e^{\lambda}_i(\vec{k})
\sigma^{\alpha}_{\lambda}(\vec{k})\quad.
\end{eqnarray*}
It is also convenient to define
\begin{eqnarray*}
& & \sigma_+^{\alpha}(\vec{k})=\sigma_L^{\alpha}(\vec{k})-i\sigma_R^{\alpha}(\vec{k})\\
& & \sigma_-^{\alpha}(\vec{k})=\sigma_R^{\alpha}(\vec{k})-i\sigma_L^{\alpha}(\vec{k})\\
& & b_+^{\alpha}(\vec{k})=b_L^{\alpha}(\vec{k})-ib_R^{\alpha}(\vec{k})\\
& & b_-^{\alpha}(\vec{k})=b_R^{\alpha}(\vec{k})-ib_L^{\alpha}(\vec{k})\\
& & b_+^A(\vec{k})=b_L^A(\vec{k})-ib_R^A(\vec{k})\\
& & b_-^A(\vec{k})=b_R^A(\vec{k})-ib_L^A(\vec{k})\\
& & \alpha_+^{\alpha}(\vec{k})=\alpha_L^{\alpha}(\vec{k})-i\alpha_R^{\alpha}(\vec{k})\\
& & \alpha_-^{\alpha}(\vec{k})=\alpha_R^{\alpha}(\vec{k})-i\alpha_L^{\alpha}(\vec{k})\quad.\\
\end{eqnarray*}
A useful formula used in the obtention of the symplectic structure
is
$$\int_{\mathbb{R}^3}\stackrel{1}{\omega}\wedge*\stackrel{2}{\omega}=
3!\int_{\mathbb{R}^3}\frac{d^3k}{w^2}\stackrel{1}{\beta}_{ij}(\vec{k},t)
\stackrel{2}b_{ij}(-\vec{k},t)+3\int_{\mathbb{R}^3}d^3k
\stackrel{1}{\alpha}_i(\vec{k},t)\stackrel{2}{a}_i(-\vec{k},t)$$
where $\stackrel{1}{\omega}$ and $\stackrel{2}{\omega}$ are a
2-form and a 3-form respectively with Fourier components given by
\begin{eqnarray*}
\stackrel{1}{\omega}=
\left\{\begin{array}{c}ik_{[i}\stackrel{1}{\alpha}_{j]}(\vec{k},t)+
\stackrel{1}{\beta}_{ij}(\vec{k},t)\\ \\
ik_i \stackrel{1}{a}(\vec{k},t)+\stackrel{1}{b}_i(\vec{k},t)
\end{array}\right\}\quad
\stackrel{2}{\omega}=
\left\{\begin{array}{c}ik_{[i}\stackrel{2}{\alpha}_{jk]}(\vec{k},t)\\ \\
ik_{[i}
\stackrel{2}{a}_{j]}(\vec{k},t)+\stackrel{2}{b}_{ij}(\vec{k},t)
\end{array}\right\}
\end{eqnarray*}

\section{\label{alg} Some algebraic issues}

Let us show, first, that
\begin{eqnarray*}
\ker\mathcal{P}\cap \ker\mathcal{Q}=\{0\}\nonumber
\end{eqnarray*}
In fact, if we write $\left[\begin{array}{c} x \\ y \end{array}
\right]\in \ker\mathcal{P}\cap \ker\mathcal{Q}$ ($x,\,y\in
\mathbb{C}^N$) we have
$$\mathcal{P}
\left[\begin{array}{c} x \\ y \end{array}
\right]=\left[\begin{array}{c} 0 \\ 0 \end{array}
\right]\Leftrightarrow\left\{\begin{array}{c} ax+by=0 \\
\overline{b}x+\overline{a}y=0\end{array} \right.\quad\mathcal{Q}
\left[\begin{array}{c} x \\ y \end{array}
\right]=\left[\begin{array}{c} 0 \\ 0 \end{array}
\right]\Leftrightarrow\left\{\begin{array}{c} ax-by=0 \\
-\overline{b}x+\overline{a}y=0\end{array} \right.$$ from which it
is immediate to get
$$(P+iR)x=0,\,(Q+iR^t)x=0,\quad(P-iR)y=0,\,(Q-iR^t)y=0.$$
i.e.
$$ x\in \ker(P+iR)\cap \ker(Q+iR^t),\quad y\in \ker(P-iR)
\cap \ker (Q-iR^t)$$ but the obvious isomorphism between
$\Omega_2^{2N}$ at each point $p$ of $ \mathcal{M}$ and
$\mathbb{C}^{2N}$ (defined by identifying * and $i$) implies that
the condition (\ref{010}) is equivalent to $\ker(P+iR)\cap \ker
(Q+iR^t)=\{0\}$ (and also to its conjugate $\ker(P-iR)\cap \ker
(Q-iR^t)=\{0\}$) so that we get $x=0$ and $y=0$ and (\ref{011}).

We prove now every vector in a linear basis of $\ker\mathcal{P}$
can be written in the form $\left[\begin{array}{c} v \\
\overline{v}
\end{array} \right]$ (for some $v\in\mathbb{C}^N$); indeed,
$$\mathcal{P}\left[\begin{array}{c} e_+ \\ e_- \end{array}
\right]=\left[\begin{array}{c} 0 \\ 0 \end{array}
\right]\Leftrightarrow \left\{\begin{array}{c} ae_+ +be_-=0\\
\overline{b}e_++\overline{a}e_-=0
\end{array} \right.\Leftrightarrow
\left\{\begin{array}{c} a\overline{e}_-+b\overline{e}_+=0\\
\overline{b}\overline{e}_-+\overline{a}\overline{e}_+=0\end{array}
\right.
\Leftrightarrow\mathcal{P}\left[\begin{array}{c} \overline{e}_- \\
\overline{e}_+
\end{array} \right]=\left[\begin{array}{c} 0 \\ 0 \end{array}
\right]$$ and, hence,
$$\left[\begin{array}{c} e_+ \\ e_- \end{array}
\right]\in \ker\mathcal{P} \Leftrightarrow \left[\begin{array}{c}
\overline{e}_- \\ \overline{e}_+\end{array} \right]\in
\ker\mathcal{P}$$ There are two possibilities now: if both
$\left[\begin{array}{c} e_+ \\ e_- \end{array} \right]$ and
$\left[\begin{array}{c} \overline{e}_- \\
\overline{e}_+\end{array} \right]$ are linearly independent then
both $\left[\begin{array}{c} e_+ \\ e_- \end{array} \right]+
\left[\begin{array}{c} \overline{e}_- \\
\overline{e}_+\end{array} \right]$ and $i\left[\begin{array}{c}
e_+ \\ e_- \end{array} \right]
-i\left[\begin{array}{c} \overline{e}_- \\
\overline{e}_+\end{array} \right]$ are, also, linearly independent
and have the form $\left[\begin{array}{c} v \\ \overline{v}
\end{array} \right]$. If, instead, $\left[\begin{array}{c} e_+
\\e_-\end{array} \right]$ and
$\left[\begin{array}{c} \overline{e}_- \\
\overline{e}_+\end{array} \right]$ are linearly dependent (and
non-zero) then there exist $\rho\in\mathbb{R},\,\rho\geq 0$ and
$\theta\in[0,2\pi)$ such that $\left[\begin{array}{c} e_+ \\ e_-
\end{array} \right]=\rho e^{i\theta}\left[\begin{array}{c}
\overline{e}_-\\ \overline{e}_+\end{array} \right]$; which gives
$e_+=\rho e^{i\theta}\overline{e}_-$ and $e_-=\rho
e^{i\theta}\overline{e}_+$ and, hence, $\rho=1$. We see that is
then possible to write
$$\left[\begin{array}{c} e_+\\e_-\end{array} \right]=
e^{i\theta}\left[\begin{array}{c}\overline{e}_-\\
\overline{e}_+\end{array} \right]=\left[\begin{array}{c}
e_+\\e^{i\theta}\overline{e}_+\end{array} \right]$$ which has the
desired form after multiplying by $e^{-i\theta/2}$. Is is a simple
matter now to build a basis with vectors in the form introduced
above by using this procedure. Given a basis of $\ker\mathcal{P}$
we can adjoin to it all the vectors obtained by switching the plus
and minus components and conjugating; substitute every vector by
one of the form $\left[\begin{array}{c} v\\
\overline{v}\end{array} \right]$ and then choose a maximally
independent set among the resulting vectors to have a basis of
vectors of this type.

\section{\label{matrices} Some useful matrix identities}

Writing
$$\mathbf{v}_+\equiv\left[\begin{array}{c} v_A \\
\overline{v}_A
\end{array} \right]\, \quad \mathbf{v}_-\equiv\left[\begin{array}{c} v_A \\
-\overline{v}_A
\end{array} \right]\, \quad \mathbf{e}_-\equiv\left[\begin{array}{c} e_{\alpha} \\
-\bar{e}_{\alpha}
\end{array} \right]\, \quad \mathbf{e}_+\equiv\left[\begin{array}{c} e_{\alpha} \\
\bar{e}_{\alpha}
\end{array} \right]$$
we have
\begin{eqnarray*}
& &\mathbf{v}_-^t\mathcal{P}\mathbf{v}_-=4\Re_{AB}\quad
\mathbf{v}_+^t\mathcal{Q}\mathbf{v}_+=4\Re_{AB} \nonumber\\
& &\mathbf{v}_-^t\mathcal{P}\mathbf{e}_+=4i\Im_{A\alpha}\quad
\mathbf{v}_+^t\mathcal{Q}\mathbf{e}_+=4\Re_{A\alpha}\nonumber\\
& &\mathbf{v}_-^t\mathcal{P}\mathbf{e}_-=4\Re_{A\alpha}\quad
\mathbf{v}_+^t\mathcal{Q}\mathbf{e}_-=4i\Im_{A\alpha}\nonumber\\
& &\mathbf{e}_+^t\mathcal{P}\mathbf{v}_-=4i\Im_{A\alpha}\quad
\mathbf{e}_+^t\mathcal{Q}\mathbf{v}_+=4\Re_{\alpha A}\nonumber\\
& &\mathbf{e}_+^t\mathcal{P}\mathbf{e}_+=2\Re^+_{\alpha\beta}\quad
\mathbf{e}_+^t\mathcal{Q}\mathbf{e}_+=2\Re^-_{\alpha\beta}\nonumber\\
&
&\mathbf{e}_+^t\mathcal{P}\mathbf{e}_-=2i\Im^-_{\alpha\beta}\quad
\mathbf{e}_+^t\mathcal{Q}\mathbf{e}_-=2i\Im^+_{\alpha\beta}\nonumber\\
& &\mathbf{e}_-^t\mathcal{P}\mathbf{v}_-=4\Re_{\alpha A}\quad
\mathbf{e}_-^t\mathcal{Q}\mathbf{v}_+=4i\Im_{\alpha A}\nonumber\\
&
&\mathbf{e}_-^t\mathcal{P}\mathbf{e}_+=2i\Im^+_{\alpha\beta}\quad
\mathbf{e}_-^t\mathcal{Q}\mathbf{e}_+=2i\Im^-_{\alpha\beta}\nonumber\\
& &\mathbf{e}_-^t\mathcal{P}\mathbf{e}_-=2\Re^-_{\alpha\beta}\quad
\mathbf{e}_-^t\mathcal{Q}\mathbf{e}_-=2\Re^+_{\alpha\beta}\nonumber
\end{eqnarray*}
It is useful to notice that
$\Im^b_{\alpha\beta}=-\Im^b_{\beta\alpha}$ and, hence
$\Im^+_{\alpha\beta}=\Im^-_{\beta\alpha}$.

\section{\label{R=0} A summary of results for R=0}

We summarize in this section the main features of the $R=0$ case
\cite{edufersform}. The field equations derived from (\ref{001})
are
\begin{eqnarray*}
P\delta d A+Qd\delta A=0\quad.
\end{eqnarray*}
Expanding $A=A^p e_p+A^q e_q+A^r e_r$ with $\ker\,P={\rm
Span}\,\{e_p\}$, $\ker\, Q={\rm Span}\,\{e_q\}$,  and
$\mathbb{R}^N={\rm Span}\, \{e_p,e_q,e_r\}$ we can solve them by
first considering the necessary conditions
\begin{eqnarray*}
&\square dA^q=0, \quad \square \delta A^p=0,\quad \square
dA^r=0,\quad \square \delta A^r=0.
\end{eqnarray*}
Their solutions are
\begin{eqnarray*}
A^q=\gamma^q+d\Lambda^q,\quad A^p=\gamma^p+\delta \Theta^p,\quad
A^r= \gamma^r+d\delta \gamma_{{\scriptscriptstyle H\!D}}^r
\end{eqnarray*} with $\Lambda^q$, and $\Theta^p$ arbitrary and
\begin{eqnarray*}
\gamma^q=\left\{\begin{array}{c}\beta^q_{ij}(\vec{k})e^{-iwt}+\bar{\beta}^q_{ij}(-\vec{k})e^{iwt}\\
\\b^q_{i}(\vec{k})e^{-iwt}+\bar{b}^q_{i}(-\vec{k})e^{iwt}\end{array}\right\}\quad,\label{023}
\end{eqnarray*}
\begin{eqnarray*}
\gamma^p=\left\{\begin{array}{c}\frac{i}{w}k_{[i}\alpha^p_{j]}(\vec{k})e^{-iwt}
+\frac{i}{w}k_{[i}\bar{\alpha}^p_{j]}(-\vec{k})e^{iwt} \\
\\\frac{i}{w}k_{i}\left[a^p(\vec{k})e^{-iwt}+\bar{a}^p(-\vec{k})e^{iwt}\right]
\end{array}\right\}\quad,
\end{eqnarray*}
\begin{eqnarray*}
\gamma^r&=&\left\{\begin{array}{c}
\beta^r_{ij}(\vec{k})e^{-iwt}+\bar{\beta}^r_{ij}(-\vec{k})e^{iwt}\\ \\
\frac{i}{w}k_{i}\left[ a^r(\vec{k})e^{-iwt}
    +\bar{a}^r(-\vec{k})e^{iwt}\right]+b^r_{i}(\vec{k})e^{-iwt}
    +\bar{b}^r_{i}(-\vec{k})e^{iwt}
\end{array}\right\}\quad,
\end{eqnarray*}
\begin{eqnarray*}
d\delta\gamma^r_{{\scriptscriptstyle H\!D}}&=&
\left\{\begin{array}{c}
  \frac{i}{w}k_{[i}\left\{
    \left[\alpha^r_{j]}(\vec{k})+wt\sigma^r_{j]}
    (\vec{k})\right]e^{-iwt}
    +\left[\bar{\alpha}^r_{j]}(-\vec{k})+wt\bar{\sigma}^r_{j]}
 (-\vec{k})\right]e^{iwt}
 \right\}\\ \\
\frac{1}{2}\left[\left(\sigma^r_{i}(\vec{k})
    -i\alpha^r_{i}(\vec{k})-iwt\sigma^r_{i}(\vec{k})\right)e^{-iwt}+
    \left(\bar{\sigma}^r_{i}(-\vec{k})
    +i\bar{\alpha}^r_{i}(-\vec{k})+iwt\bar{\sigma}^r_{i}(-\vec{k})\right)e^{iwt}\right]
\end{array}\right\}.
\end{eqnarray*}
The objects appearing in the previous expressions are subject to
the algebraic conditions
\begin{eqnarray*}
P\left[b^q_{i}(\vec{k})e_q+b^r_{i}(\vec{k})e_r\right]=Q\left[
\frac{i}{2}\alpha^p_{i}(\vec{k})e_p+b^r_{i}(\vec{k})e_r+
\sigma^r_{i}(\vec{k})e_r\right]\, .
\end{eqnarray*}
The symplectic structure is given by
\begin{eqnarray*}
\Omega=\int_{\mathbb{R}^3}\frac{d^3\vec{k}}{w}\bigg\{-4i\dd
\bar{\beta}^t_{ij}(\vec{k})\ww P\dd \beta^{ij}(\vec{k}) +8i\dd
\bar{a}^t(\vec{k})\ww Q \dd a(\vec{k})+8i\dd
\bar{b}_{i}^r(\vec{k})e_r^t\ww P\dd b_{i}^r(\vec{k})
& &\nonumber\\
+4i\dd \bar{b}_{i}^r(\vec{k})e_r^t\ww P\dd b_{i}^q(\vec{k})+4i\dd
\bar{b}_{i}^q(\vec{k})e_r^t\ww
P\dd b_{i}^r(\vec{k})& &\nonumber\\
+2i\left[\dd \bar{\sigma}^r_{i}(\vec{k}) e^t_r\ww P\dd
b^{i}(\vec{k})- \dd \sigma^r_{i}(\vec{k}) e^t_r\ww P\dd
\bar{b}^{i}(\vec{k}) \right]& &\nonumber\\
-4\left[ \dd\bar{\alpha}^r_{i}(\vec{k})e^t_r\ww P\dd
b^{i}(\vec{k}) + \dd \alpha^r_{i}(\vec{k})e^t_r\ww P\dd
\bar{b}^{i}(\vec{k}) \right]\bigg\},& &\nonumber
\end{eqnarray*}
the energy momentum is
\begin{eqnarray*}
\tau^a\dd P_a=\dd\bigg\{
\int_{\mathbb{R}^3}\frac{d^3\vec{k}}{w}\bigg[(\tau^ak_a)[ -4
\bar{\beta}^t_{ij}(\vec{k})P \beta_{ij}(\vec{k}) +8
\bar{a}^t(\vec{k})Q a(\vec{k})+8\bar{b}^r_{i}(\vec{k})
e_r^t P b^r_i(\vec{k})+& & \nonumber\\
4\bar{b}^r_{i}(\vec{k}) e_r^t P b^q_i(\vec{k})+
4\bar{b}^q_{i}(\vec{k}) e_r^t P b^r_i(\vec{k})+
4i\bar{\alpha}^r_{i}(\vec{k})e^t_rP
b^{i}(\vec{k})-4i\alpha^r_{i}(\vec{k})e^t_rP\bar{b}^{i}(\vec{k})]&
&\nonumber\\ +2(\tau^ak_a-\tau^0
w)[\bar{\sigma}^r_{i}(\vec{k})e^t_rP
b^{i}(\vec{k})+\sigma^r_{i}(\vec{k})e^t_rP
\bar{b}^{i}(\vec{k})]\bigg]\bigg\},\nonumber
\end{eqnarray*}
and the helicity density
\begin{eqnarray}
\lambda(\vec{k})=\epsilon_{ijk}\frac{k^i}{w}\left\{
-8i\bar{b}^r_j(\vec{k})e_r^{t}Pe_{r^{\prime}}b_k^{r^{\prime}}(\vec{k})
-4i\bar{b}^r_j(\vec{k})e_r^{t}Pe_qb_k^q(\vec{k})
-4i\bar{b}^q_j(\vec{k})e_q^{t}Pe_rb_k^r(\vec{k}) \right.& &
\nonumber\\
\left.-2i[\bar{\sigma}^r_j(\vec{k})+2i\bar{\alpha}^r_j(\vec{k})]Pb_k(\vec{k})+
2i[\sigma^r_j(\vec{k})-2i\alpha^r_j(\vec{k})]P\bar{b}_k(\vec{k})\right\}.&
&  \nonumber
\end{eqnarray}
In all these expressions we have $ \beta_{ij}\equiv \beta^r_{ij}
e_r +\beta_{ij}^q e_q $, $a\equiv a^pe_p+a^r e_r $, and $b_i\equiv
b_i^r e_r +b_i^q e_q $.

\begin{acknowledgments}
The authors wish to thank J. Le\'on and A. Tiemblo for several
interesting discussions. E.J.S.V. is supported by a Spanish
Ministry of Education and Culture fellowship co-financed by the
European Social Fund.
\end{acknowledgments}


\end{document}